\documentclass[preprint,12pt]{aastex}  
\usepackage{epsfig}

\def\Ka{K$\alpha$~}

\def\fexxv{Fe\,{\sc xxv}}
\def\fexxvi{Fe\,{\sc xxvi}}
\def\civ{C\,{\sc iv}}
\def\oi{O\,{\sc i}}

\def\nvii{N\,{\sc vii}}
\def\nevii{Ne\,{\sc vii}}
\def\nex{Ne\,{\sc x}}

\def\mgv{Mg\,{\sc v}}
\def\mgxii{Mg\,{\sc xii}}
\def\naxi{Na\,{\sc xi}}
\def\sivi{Si\,{\sc vi}}
\def\sixiv{Si\,{\sc xiv}}
\def\nixix{Ni\,{\sc xix}}
\def\aa{\buildrel_{\circ}\over{\mathrm{A}}}

\def\fexxv{Fe\,{\sc xxv}}
\def\fexxvi{Fe\,{\sc xxvi}}
\def\civ{C\,{\sc iv}}

\def\mathbi#1{\textbf{\em #1}}
\def\mathv{\textbf{\em v}}
\def\mathB{\textbf{\em B}}

\def\cm{\ifmmode {\rm cm}^{-1} \else cm$^{-1}$ \fi}
\def\s{\ifmmode {\rm s}^{-1} \else s$^{-1}$ \fi}
\def\cc{\ifmmode {\rm cm}^{-3} \else cm$^{-3}$ \fi}
\def\cs{\ifmmode {\rm cm}^{-2} \else cm$^{-2}$ \fi}
\def\g{\ifmmode \gamma \else $\gamma$\fi}
\def\G{\ifmmode \Gamma \else $\Gamma$\fi}
\def\Gs{\ifmmode \Gamma~ \else $\Gamma~$\fi}
\def\Ka{K$\alpha$~}

\def\aoxx{$\alpha_{OX}~$}
\def\gc{\ifmmode \gamma_{\rm c} \else $\gamma_{\rm c}$ \fi}
\def\sw{Schwarzschild~}
\def\gsim{\mathrel{\raise.5ex\hbox{$>$}\mkern-14mu
             \lower0.6ex\hbox{$\sim$}}}
\def\lsim{\mathrel{\raise.3ex\hbox{$<$}\mkern-14mu
             \lower0.6ex\hbox{$\sim$}}}
\def\simless{\mathbin{\lower 3pt\hbox
     {$\rlap{\raise 5pt\hbox{$\char'074$}}\mathchar"7218$}}}   
\def\simmore{\mathbin{\lower 3pt\hbox
     {$\rlap{\raise 5pt\hbox{$\char'076$}}\mathchar"7218$}}}   
\def\Msun{M_\odot}                                
\def\4u{4U 1728--34}
\def\deg{^\circ}
\def\aa{\buildrel _{\circ} \over {\mathrm{A}}}

\lefthead{et al.} \righthead{\4u}

\shorttitle{High-Velocity Outflows in BAL QSOs}

\shortauthors{Kazanas et al.}

\begin{document}

\title{Toward a Unified AGN Structure }


%
\author{\textsc{Demosthenes Kazanas}\altaffilmark{1,2},
\textsc{Keigo Fukumura}\altaffilmark{2,3},
\textsc{Ehud Behar}\altaffilmark{4}, \\
\textsc{Ioannis Contopoulos}\altaffilmark{5}\\
\textsc{and} \\
\textsc{Chris Shrader}\altaffilmark{2,6} }

\altaffiltext{1}{Email: Demos.Kazanas@nasa.gov}
\altaffiltext{2}{Astrophysics Science Division, NASA/Goddard Space Flight
Center, Greenbelt, MD 20771}
\altaffiltext{3}{University of Maryland, Baltimore County
(UMBC/CRESST), Baltimore, MD 21250}
 \altaffiltext{4}{Research Center for Astronomy, Academy of
Athens, Athens 11527, Greece} \altaffiltext{5}{Department of
Physics, Technion, Haifa 32000, Israel}\altaffiltext{6}{Universities
Space Research Association}

\begin{abstract}

\baselineskip=15pt

We present a unified model for the structure and appearance of
accretion powered sources across their entire luminosity range from
galactic X-ray binaries  to  luminous quasars, with emphasis on AGN
and their phenomenology. Central to this model is the notion of MHD
winds launched from the accretion disks that power these objects.
These winds provide the matter that manifests as blueshifted
absorption features in the UV and X-ray spectra of a large fraction
of these sources; furthermore, their density distribution in the
poloidal plane determines the ``appearance" (i.e. the column and
velocity structure of these absorption features) as a function of
the observer inclination angle. This work focuses on just the
broadest characteristics of these objects; nonetheless, it provides
scaling laws that allow one to reproduce within this model the
properties of objects spanning a very wide luminosity range and
viewed at different inclination angles, and trace them to a common
underlying dynamical structure. Its general conclusion is that the
AGN phenomenology can be accounted for in terms of three parameters:
The wind mass flux in units of the Eddington value, $\dot m$, the
observer's inclination angle $\theta$ and the logarithmic slope
between the O/UV and X-ray fluxes $\alpha_{OX}$. However, because of
a significant correlation between \aoxx and UV luminosity, we
conclude that the AGN structure depends on only two parameters.
Interestingly, the correlations implied by this model appear to
extend to and consistent with the characteristics of galactic X-ray
sources, suggesting the presence of a truly unified underlying
structure for accretion powered sources.

\end{abstract}


\baselineskip=15pt

\section{Introduction}

The notion of AGN as an astronomical object of solar system
dimensions and luminosity surpassing that of a galaxy has been with
us for about half a century now. Since then, the advent of novel
observational techniques, the accumulation of data and theoretical
modeling has refined and advanced our notions as to what constitutes
an AGN, with accretion onto a black hole as the source of the
observed radiation now being universally accepted. At the same time,
the discovery of galactic bright X-ray binary (XRB) sources, powered
also by accretion onto compact objects (neutron stars and stellar
size black holes) has extended the notion of accretion powered
source to the stellar domain. Indeed, the general similarity of the
X-ray spectral properties of AGN and galactic black hole candidates
(GBHC) and XRBs in general, including their broad Fe \Ka
fluorescence features \citep{Miller07}, argues for near horizon
structures which are very similar, despite the huge disparity in the
objects' scales. This structure is thought to consist of a
Shakura-Sunyaev \citep{SS73} disk that extends to the ISCO
(innermost stable circular orbit) of the corresponding flow,
supplemented by an overlying hot, X-ray emitting corona.

Even though it is generally accepted that the AGN radiant energy is
released by the accretion of matter onto a black hole (or in certain
cases by extraction of the hole's rotational energy) in a region
comparable to its horizon, there is plenty of evidence that a
significant fraction of the AGN power is emitted, after
reprocessing, at much larger radii.
[One should note however, that accretion energy can also be
transported outward not only radiatively but also mechanically by
the viscous stresses that transport the accretion flow's angular
momentum \citep{BB99}].
%
Thus, the UV and optical lines that constitute, typically, a
fraction $f \sim 10\%$ of the AGN bolometric luminosity, are emitted
presumably by clouds at distances $\sim 0.1 - 10$ pc that cover a
fraction $f$ of the AGN solid angle. In addition to the line
emission, the AGN ionizing continuum is also reprocessed into IR and
far--IR radiation by matter at even larger distances, which
apparently subtends an even larger fraction of the AGN solid angle
($\sim 50\%$).
%
The geometry of this component is thought to be cylindrical (rather
than spherical) with a column density that depends strongly on the
angle $\theta$ of the observers' line of sight (LoS) with the
symmetry axis. It was proposed that such a geometry nicely unifies
the Seyfert-1 and Seyfert-2 AGN subclasses \citep{AntMil85} and also
those of the broad and narrow line radio galaxies (BLRG - NLRG)
\citep{Barthel89}, according to the angle $\theta$: Thus, Seyfert-1s
(or BLRG) are AGN in which the observer's LoS makes a small angle
with their axis of symmetry, the column of the intervening cold gas
is small ($N_H < 10^{21} {\rm cm}^{-2}$) and the continuum source
and its surrounding broad line emission (concentrated in the inner
AGN regions) are directly visible. Seyfert-2s (or NLRG) on the other
hand, represent the same objects viewed at a large inclination
angle, along which the column density to the source is much larger
($N_H > 10^{23} {\rm cm}^{-2}$), obscuring the continuum source and
allowing the view of only the large distance (hence narrow
component) of the emission lines. This obscuring structure is
referred to as the ``AGN molecular torus", considering that it must
consist of gas in molecular state, given its low effective
temperature ($T \sim 10 - 100$ K). Statistics of Seyfert-1 and
Seyfert-2 AGN imply that the height $h$ of these torii must be
comparable to their distance $R$ from the AGN center, i.e. $h/R
\simeq 1$. However, the value of this ratio is in conflict with that
implied by hydrostatic equilibrium and the ratio of their thermal
($v_{\rm th} \sim 1 \, (T/ {\rm 100 K})$ km/s) and Keplerian
($v_{\rm K} \sim 300 - 500$ km/s) velocities, namely $h/R \simeq
v_{\rm th}/v_{\rm K} \sim 10^{-3}$, thus presenting us with a
conundrum concerning the physics of these structures.

These spectroscopically inferred components, along with observations
of narrow radio jets along the AGN symmetry axis, led to the now
well known AGN picture of \citep{UP95}, which consists simply of
their arrangement  at the appropriate positions in the AGN vicinity.
Compelling as this picture might be observationally, it includes
very little, if any, of the underlying physics. The AGN constituent
components are independent of each other with physical properties
assigned as needed by the observations of the specific objects.
However, more recent observational developments suggest that such a
picture is rather incomplete. To begin with, \citet{BG92} have shown
the existence of interrelations among AGN the line properties and
also relations to other bands of the spectrum (notably the X-rays).
Then, the increase in UV spectral resolution afforded by {\em HST}
has shown that roughly 50\% of Seyfert-1s exhibit UV absorption
troughs due to plasma outflowing at $v \simeq 300-1000$ km/s, too
narrow to have been discerned by the earlier {\em IUE} observations
\citep{Cren99}, which did detect some, but in a much smaller
fraction of the overall AGN population. To these flows one must also
include those of the so-called BAL QSOs, which reach velocities
along the observer's LoS in excess of $10^4$ km/s \citep{Weymann91}.
These are observed in about $\simeq 10\%$ of high luminosity
quasars, implying that they subtend a similar fraction of the
continuum source solid angle in these objects.

In addition to these UV absorption features, outflowing components
were also found in the AGN X-ray spectra. The increase in spectral
resolution provided by {\em ASCA} showed that approximately $50\%$
of Seyfert-1s exhibit also blue-shifted absorption features in their
X-ray spectra \citep{George00}, indicative of outflowing plasma, but
of different ionization state than that responsible for the UV
absorption features. More recently, \citet{Tombesi} have shown that
Fe-K absorption features at velocities $v \sim 0.1c$ are rather
common in nearby Seyfert galaxies and coined for them the term
ultra-fast outflows (UFO). The simultaneous presence of both UV and
X-ray absorbers in the same objects implies they belong to the same
outflowing plasma \citep[see e.g.][]{Gab03}. However, despite a
large number of studies supporting this hypothesis,
\citep[][]{Mathur94,Mathur95,Collinge01,Crenshaw03,Brandt09}, an
understanding of the underlying gas dynamics is lacking. A common
origin for the plasma of these components as features of a common,
radiatively driven flow would be hard to reconcile with their
different velocities and ionization properties.

An account of the observed AGN outflows, in particular of the most
challenging high velocity ones of BAL QSOs was put forward
semi-analytically by \citet{MCGV95}. These authors, in analogy with
the winds of O-stars, proposed that they  are driven off the inner
regions of the QSO accretion disks by UV and optical line radiation
pressure to achieve velocities consistent with those observed. The
same issue was taken up in more detail  in 2D numerical calculations
by \citet{PSK00} who included in these calculations the detailed
photoionization of the line driven wind by the QSO X-ray radiation.
As shown in this work, efficient wind driving by line pressure
requires that the line driven wind material be shielded from the
ionizing effects of the X-rays, otherwise line driving becomes
ineffective. Their calculations showed that the ``failed wind" from
the highly ionized innermost regions of the AGN accretion disk did
provide the required shielding. The fact that BAL QSOs are weak
X-ray emitters appears to advocate for such a point of view.

The ubiquity of AGN outflows implies that they should be included in
the AGN structure schematic of \citet{UP95}; however, the broad
range of observed velocities and their different values in the UV
and X-ray bands make such a construct complicated in the absences of
an underlying unifying principle. However, such attempts have been
made. Thus, \citet{Elvis00}, motivated by the velocity fields
produced by \citet{PSK00} in modeling the BAL QSO outflows, proposed
a scheme that would supplement the AGN picture of \citet{UP95} with
outflow components consistent with observed phenomenology. By
limiting the fast ($v \gsim 10^4$ km/s), radiatively driven flow to
a narrow angular sliver ($\Delta \theta \simeq 6^{\circ}$) around
$\theta_s \simeq 50^{\circ}$ \citep{PSK00}, he accounted for the
observed fraction of BAL QSOs in the overall QSO population. He then
attributed the lower velocities of the typical X-ray and UV
absorption features to the projection effects of viewing this flow
at a larger angle $(\theta > \theta_s)$ and the absence of
absorption features in fraction of the objects to the low column and
high ionization of the wind at $\theta < \theta_s$. He also
postulated that the angular position $\theta_s$ and the opening
angle $\Delta \theta$ of the high velocity radiatively driven radial
stream would vary with source luminosity in a way that could account
the variation of source properties with luminosity. However, his
approach ignored the outflows seen in the AGN X-ray spectra, which
at the time were not as well documented.

The AGN picture proposed herein is in the same spirit as that of
\citet{Elvis00} in that, employing a well defined wind dynamical
model, it provides a framework of systematizing the multitude of
observational facts, in particular the more recent high resolution
X-ray spectroscopy observations. However, it does more than that; it
provides, in addition, scaling arguments similar to those put
forward by \citet{Boroson02}, which allow one to incorporate within
this single framework the ionization properties of Seyferts, BAL
QSOs and XRBs, the structure of the AGN molecular torii and the
corresponding IR spectra. This is possible because the underlying
dynamical models, which span many decades in radius, are to a large
extent independent of the mass of the accreting object and, as such,
they can be applied to objects over a very wide range of luminosity.
As we will discuss in the ensuing sections, the models we present
provide the possibility of a broader classification of the structure
of accreting sources in terms of a small number of parameters (2)
thus providing an opportunity for a unified treatment of all
accretion powered sources. {  The present work will concentrate on
the structure of AGN, however it will be argued that the structures
of XRBs are quite similar, their appearance being different only
because of their very different ionization.}


In \S 2 we provide a brief review of AGN outflow phenomenology with
emphasis on the more recent high spectral resolution observations of
X-ray absorbers by {\em Chandra} and {\em XMM-Newton}. In \S 3 we
present our model and its general scalings. In \S 4 the model is
applied to produce the absorber properties of galactic and
extragalactic objects as specific cases of its parameters along with
its general structure properties, depicted in two diagrams that
relate the absorber column, velocity and observation angle. In \S 5
we focus on the emission properties of these winds and provide an
account of the observed linear relation between the H$\alpha$ and
bolometric AGN luminosities as well as their   IR--to--far-IR
spectra. Finally in \S 6 the results are reviewed and directions for
future research are outlined.

\section{X-Ray Spectroscopy and Warm Absorbers}

Following the discovery of the ubiquitous nature of X-ray absorption
features (Warm Absorbers) in the {\em ASCA} spectra, the launch of
{\em Chandra} and {\em XMM-Newton} ushered a new era in the study of
these features. Their superior sensitivity and resolution compared
with those of previous missions made clear the presence of a
plethora of transitions \citep[see e.g.][]{Behar03} in the spectra
of numerous AGN. For example, the X-ray bright QSO IRAS~13349+2438
($z=0.10764$) and the Seyfert-1 galaxy (e.g. MCG-6-30-15 at
$z=0.007749$) observed extensively with {\it ROSAT}
\citep[][]{Brandt96} and {\it ASCA} \citep[][]{Brandt97} gave
indications of absorption features in the sub-keV regime. Their {\em
Chandra} and {\em XMM-Newton} spectra confirmed this fact and
exhibited a wealth of transitions such as \nvii, \nevii--\nex,
\mgv--\mgxii, \naxi, \sivi--\sixiv, and \nixix~ while successfully
identifying almost all charge states of Fe and O in some cases
\citep[][hereafter HBK07]{Behar03,HBK07}. Furthermore, these
observations showed the various transitions to be blueshifted
relative to the host galaxy, indicating that the warm absorber
plasma is, in fact, outflowing.

Though the majority of the X-ray absorber features in the {\em
Chandra} and {\em XMM-Newton} archives are associated with the
spectra of Seyfert galaxies, absorption features have been also seen
in the spectra of quasars; these are mainly Fe-K features usually at
velocities higher than those seen in Seyferts, in the range of $v/c
\sim 0.1 - 0.8$, suggesting launching of these winds from regions
very close to the compact object. These include both BAL [see
\cite{Chartas03} for PG~1115+080, \cite{Chartas07} for H~1413+117,
\cite{Chartas02,Chartas09} for APM~08279+5255; as such these also
exhibit high velocity UV absorption features] as well as non-BAL
quasars [see \cite{Pounds03} for PG~0844+349, \cite{Reeves03} for
PDS~456, \cite{PP06} for PG~1211+143]. The columns of these features
are $N_H \sim 10^{23}-10^{24}$ cm$^{-2}$ which are typically at
least $\sim 10$ times higher than that of UV absorbers
\citep{SP00,Chartas09}. Finally, as discussed above, Fe-K features
at velocities $v > 10^4$ km/s, have been detected in a large
fraction ($\sim 1/3$) of observed AGN.  \cite{Tombesi} refer to
these as ultra-fast outflows (UFO).

Absorption features of varying ionization states have also been
observed in galactic sources (GBHC, XRB)
\citep[e.g.][]{Miller08,NRL,BS00}. Considering that the presence of
the ionic species observed  in these sources, like in AGN, is due to
photoionization of an outflow by the photons of the continuum
source, their distribution is determined by the photoionization
parameter $\xi = L/nr^2$, where $L$ is the ionizing luminosity, $n$
the local gas density and $r$ the distance of this gas from the
source. The higher S/N ratios of galactic sources afforded better
determination of the velocities of the different ions as well as the
densities of the outflowing wind at a given radius $r$. Based on
these parameters, it was concluded (see e.g.  \S 3.1) that the wind
of GRO 1655-40 \citep{Miller08} cannot be driven by the radiation
pressure or by the X-ray heating of this source and therefore it
must be driven by the action of magnetic fields
\citep[e.g.][]{BP82,CL94}. Similar analysis of the X-ray spectra of
GRS 1915+105 by \citet{NRL} concluded that the wind mass flux at the
outer edge of its accretion disk (associated with the lowest
ionization ions) could be as much as twenty times larger than the
mass flux needed to power the source's X-ray luminosity, while a
similar conclusion was reached by \citet{Behar03} concerning the
wind of NGC 3783. Finally, it should be noted here that the wind
velocities of the highest ionization species (Fe-K) are
significantly smaller in GBHC than in Seyferts,  which in turn are
smaller than those of BAL QSOs, a fact that should be accounted
within the general framework of any wind model.

\subsection{The Absorption Measure Distribution}


The very detection in the AGN X-ray spectra of species of such
diverse states of ionization as \fexxv, \mgv~and \oi~is an important
fact in itself. It implies a wind with density distribution that
produces ionic columns, sufficiently large to be detected, for ions
that ``live'' at widely different values of $\xi$ and, most likely,
also at widely different distances from the ionizing source. This
alone is a significant constraint on outflow models. Taking this
argument into consideration, \citet{HBK07} and subsequently
\citet{B09} fit the ensemble of the absorber data in the X-ray
spectra of a number of AGN with a continuous distribution in $\xi$
(rather than adding components of different $\xi$ until a
sufficiently low $\chi^2$ is achieved). In doing so, they developed
a statistical measure of the plethora of the transitions in the {\em
Chandra/XMM-Newton} X-ray spectra, the absorption measure
distribution (AMD). This is the hydrogen equivalent column density
of specific ions, $N_H$, per decade of ionization parameter $\xi$,
as a function of $\xi$, i.e. $AMD(\xi) = d N_{\rm H}/d \log\xi$. For
a monotonic distribution of the wind density $n(r)$ with the radius
$r$, determination of AMD is tantamount to determining the ionized
wind's density dependence on $r$ (along the observers' LoS).
Therefore, considering that $AMD \propto N_H$, a power law
dependence on $\xi$ of the form $AMD \propto N_H \propto
\xi^{\alpha}$, implies also a power law dependence for the wind
density on $r$  i.e.  $n(r) \propto r^{-s}$ with $s = (2 \alpha
+1)/( \alpha +1)$. So for $\alpha \simeq 0$ ($AMD$ independent of
$\xi$), $s \simeq 1$, consistent with the conclusion reached by
\citet{B09}.

As noted in \citet{B09}, an $n(r) \propto 1/r$ density profile is
inconsistent with the asymptotic dependence of a mass conserving
spherical wind which obtains $n(r) \propto r^{-2}$ and whose
ionization parameter ``freezes" to a constant value $\xi_{\infty}$.
On the other hand, considering that the velocity of such winds is
achieved only asymptotically, with the velocity having the general
behavior $v(r) = v_{\infty}(1-r_*/r)$ \citep{MCGV95}, the $AMD$
would exhibit in this case too a dependence on $\xi$ at radii $r
\lsim r_*$. Assuming a mass conserving spherical wind with the above
velocity structure, i.e. $n(r) v(r) r^2 = \dot m = const.$, one can
easily deduce that
\begin{equation}
\frac{dN_H}{d{\rm log} \, \xi} \sim N_H(\xi) = \frac{\dot
m}{v_{\infty} r_*} \left[ \frac{\xi_{\infty}}{\xi}-1 \right]
\end{equation}
where $\xi_{\infty}$ is the ionization parameter at infinity. Under
the above velocity field, near $r_*$, the wind base, $\xi$ (and also
$v$) approaches zero and the gas column diverges so that $AMD
\propto N_H \simeq (\dot m/v_{\infty} r_*)(\xi_{\infty}/{\xi})$. At
large distances, with $\xi = \xi_{\infty} -
\epsilon,~(\epsilon/\xi_{\infty} \ll1$), $N_H = (\dot m/v_{\infty}
r_*) \, \epsilon$, so that the absorption measure distribution has
the form $AMD \propto (\epsilon/\xi_{\infty})\propto  r_*/r$, i.e.
the wind column decreases like $r^{-1}$, as expected. The arguments
leading to the above scalings are certainly overly simple, however
they show that for radiatively driven winds, the column {\em
decreases} with {\em increasing} ionization and {\em increasing}
velocity, in significant disagreement with the dependence found by
\citet{HBK07} and \citet{B09}; they also suggest that in attempting
to account for the AMD behavior, one should search for wind models
with density and velocity profiles asymptotically different from
those driven by radiation pressure.

While it is hard to produce radiation pressure driven winds with the
density profile implied by the AMD, i.e. $n(r) \propto r^{-1}$ (even
an extended source of radiation asymptotically appears point-like
and the winds acquire density $n(r) \propto r^{-2}$), MHD winds off
accretion disks were known to provide such density profiles for some
time now \citep[][hereafter, CL94]{CL94}. These are generalizations
of the two dimensional MHD winds enunciated  by \citet{BP82} in that
they allow also for mass flux rates which depends on the radius.
Given that the velocities of these winds asymptotically scale like
the disk Keplerian velocity at the point of launching, i.e. $v
\propto r^{-1/2}$, one can then easily see that for $n(r) \propto
r^{-1}$, $\dot m (r) \propto r^2 n(r) v(r) \propto r^{1/2}$, i.e.
the wind mass flux $\dot m$ increases with distance. In this respect
these winds follow the ADIOS prescription discussed by \citet{BB99}.
The specific density profile inferred from the AMD is interesting in
that it provides equal column per decade of radius. As such, it is
consistent with the density needed to account for the lags in the
power spectra of XRB and AGN \citep{KHT,PNK01} in terms of Compton
scattering. Finally, this same profile for the dust density of AGN
winds, at radii beyond the dust sublimation radius, was employed by
\citet{KK94} and \citet{RR95} to produce Seyfert-2 and QSO
IR-spectra in agreement with observations. The fact that this wind
density profile can accommodate in a natural way several diverse and
independent aspects of AGN phenomenology under the same framework is
suggestive of its central role in the AGN structure and appearance.

\section{The Model MHD Winds}

Motivated by the straightforward interpretation of the AMD in terms
of density profiles associated with MHD winds, we outline in this
section some properties of these wind models with emphasis on their
scalings with the black hole mass, mass flux rate and distance. We
pay particular attention on the scalings of the winds' ionization
and columns, as these are related to the observations of X-ray
absorbers.

The MHD winds considered here are launched by accretion disks
threaded by poloidal magnetic fields under the combined action of
rotation, gravity and magnetic stresses as discussed in
\citet{BP82}. To simplify the treatment it is assumed that the wind
is axisymmetric and as such one need only solve the field geometry
and fluid flow in the poloidal ($r, \theta$)-plane. To simplify
further the problem one looks for solutions separable in
$r$ and $\theta$ with power law $r-$dependence for all magnetic
field and fluid variables. After consideration of all conserved
quantities in the poloidal plane (mass flux per magnetic field flux,
angular momentum, magnetic line angular velocity and Bernoulli
integral) one is left with the force balance in the
$\theta-$direction (the so-called Grad-Safranov equation). The
solution of this equation provides the angular dependence of all the
fluid and magnetic field variables with their initial values given
on the surface of the disk at ($r_o, 90\deg$), where $r_o$ is a
fiducial radius that sets the radial scale of the system and it is
of order of the Schwarzschild radius, $r_S$. The Grad-Safranov
equation has the form of a wind equation with several critical
points, the most important of which for our purposes the Alfv\'en
point; this can be crossed by appropriately choosing the flow
boundary conditions on the disk surface (see CL94, BP82).

The scalings of the magnetic field, velocity, pressure and density
are given by
\begin{eqnarray}
\mathB(r,\theta) &\equiv& (r/r_o)^{-(s+1)/2}
\tilde{\mathB}(\theta)B_o \ ,
 \label{eq:Bpol} \\
\mathv(r,\theta) &\equiv& (r/r_o)^{-1/2} \tilde{\mathv}(\theta)v_o \ ,
 \label{eq:vel} \\
p(r,\theta) &\equiv& (r/r_o)^{-(s+1)} {\cal P}(\theta)B_o^2 \ ,
 \label{eq:pres2}  \\
n(r,\theta) &\equiv& (r/r_o)^{-s} \tilde{n}(\theta)B_o^2 v_o^{-2} \
 .
 \label{eq:dens}
\end{eqnarray}
The density normalization is given in terms of the poloidal field on
the disk $B_o$, however, it is more instructive to express it in
terms of the accretion or wind outflow rate $\dot m$ as discussed in
FKCB10; then the density normalization at the inner edge of the disk
at $\theta = 90^{\circ}$ is given on  setting
$\tilde{n}(90^{\circ})=1$ by
\begin{equation}
n_o = \frac{f_W {\dot m_o}}{2\sigma_T r_S} \ ,
\label{eq:dnorm}
\end{equation}
where $f_W$ is the ratio of of the mass-outflow rate in the wind to
the mass-accretion rate $\dot{m_o}$, assumed here to be
$f_W \simeq 1$, and $\sigma_T$ is the Thomson cross-section. It is
important to note here that because the mass flux in these winds
depends in general on the radius, the normalized parameter used
throughout this work, $\dot m_o$, always refers to the mass flux at
the innermost value of the flow radius, i.e. at $r/r_o \equiv x
\simeq 1$. With the scalings given above we then have $\dot m(x) =
\dot m_o x^{-s + 3/2}$, so for $s = 1$,  as inferred from the fits
of HBK07, $\dot m \propto x^{1/2}$ (the \citet{BP82} solution has
$s=3/2$, so  $n(r) \propto x^{-3/2}$ and therefore $\dot m \propto
x^0$ or $d \dot m/ dlog x =const.$).

\begin{figure}[h]
\begin{center}$
\begin{array}{cc}
\includegraphics[trim=0in 0in 0in
0in,keepaspectratio=false,width=2.9in,angle=-0] {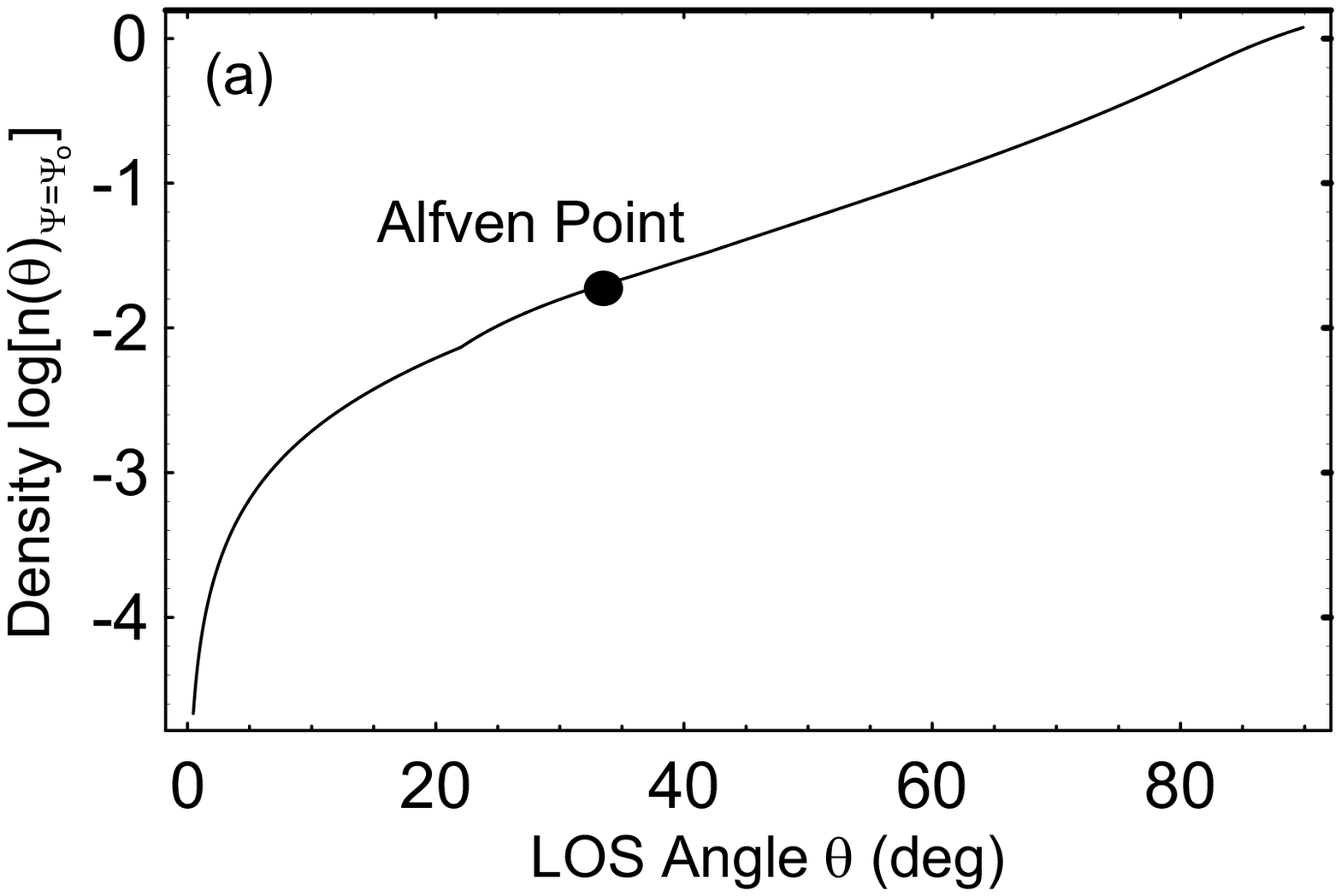} &
\includegraphics[keepaspectratio=false,width=2.9in,angle=-0]
{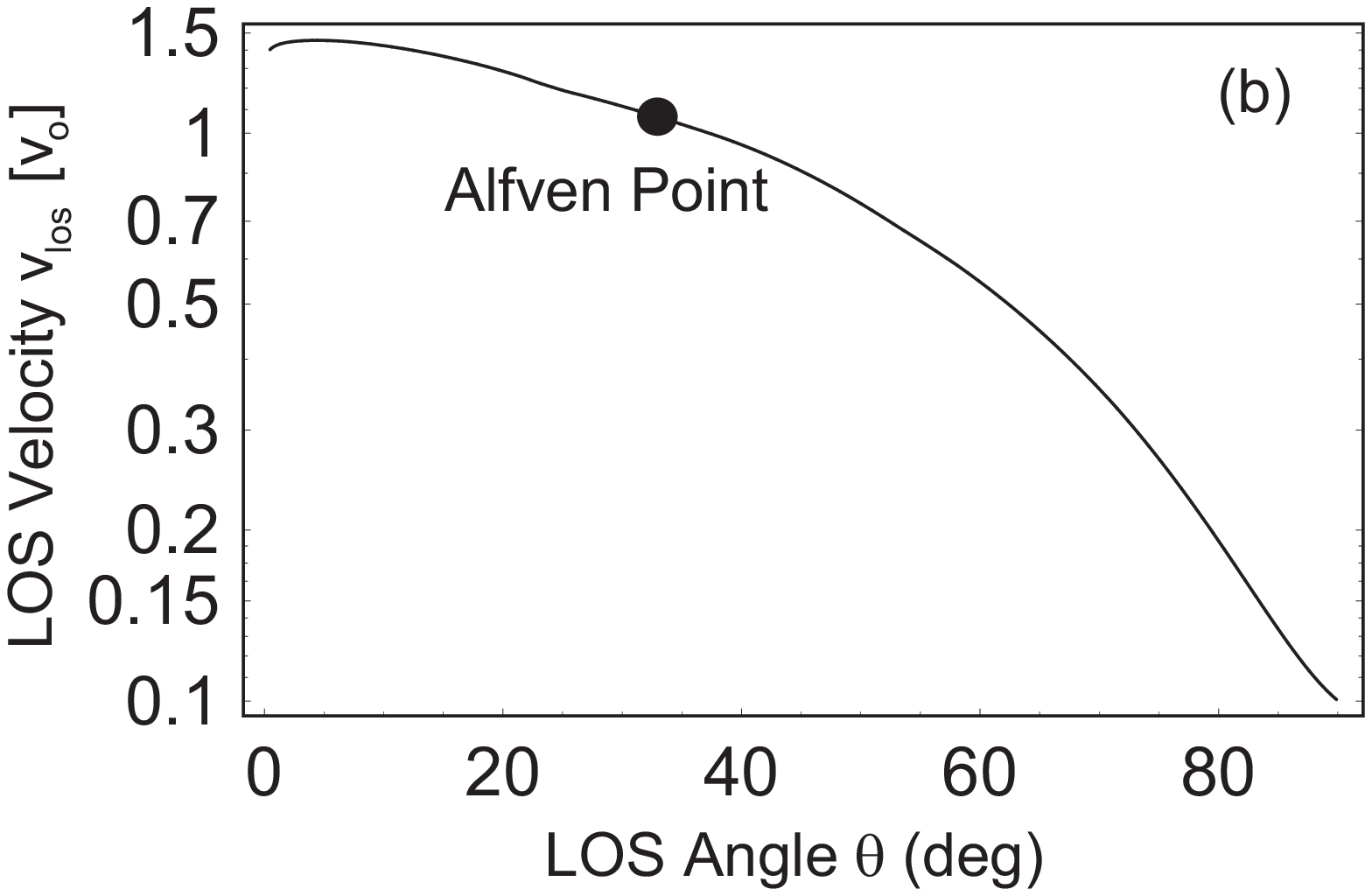}
\end{array}$
\end{center}
\caption{\small (a) Wind density as a function of the polar angle $\theta$ normalized
to its value on the disk surface ($\theta = 90^{\circ}$). (b) The line of
sight wind velocity on a given field line, normalized to the Keplerian velocity
of the foot point of the specific field line $v_o$. }
\label{fig:fig1}
\end{figure}


\subsection{Scaling Laws}

The reason that the winds we present here can be applied to AGN as
well as to GBHC, as mentioned above, is that winds and flows in
general are known to be self-similar when: The radius $r$ is
normalized to the \sw radius $r_S$ ($x = r/r_S$, $r_S= 3M$ km, where
$M= {\rm M/M}_{\odot}$ is the mass of the accreting object M in
units of the solar mass M$_{\odot}$); the mass flux $\dot M$ is
expressed in units of the Eddington accretion rate $\dot M_E =
L_E/c^2$ ($L_E$ is the Eddington luminosity, $L_E \simeq 1.3 \cdot
10^{38} \, M$ erg s$^{-1}$) as $\dot m = \dot M /\dot M_E \propto
\dot M/M$; their velocities are Keplerian, i.e. $v \simeq x^{-1/2}
\,c$.

While the wind scalings of eqs. (\ref{eq:Bpol}) - (\ref{eq:dens})
were introduced for the solution of the MHD equations  one can write
quite general scaling laws that incorporate those of the above
equations as special cases. We do so now both for accretion flows
and winds and use them to produce scalings for expressions of the
wind density, column density, ionization parameter and AMD as a
function of the dimensionless radius $x$, the dimensionless mass
flux rate $\dot m$ and black hole mass $M$.

{\bf{\em 1. Accretion:}} In applying the scalings of eqs.
(\ref{eq:Bpol}) - (\ref{eq:dens}) to spherical accretion one obtains
for the density $n(x) \propto x^{-3/2} \dot m M^{-1}$ and for the
column density $N_H(x) \propto x^{-1/2} \, \dot m$, i.e. an
expression {\em independent} of $M$; so, flows onto objects of
widely different masses but of the same $\dot m = \dot M/\dot M_E$
have the same column at the same distance $x$ from the accreting
object; its normalization is such to yield Thomson depth $\tau_T
\simeq 1$ for $x \simeq 1$ at $\dot m \simeq 1$ (this is the
similarity of Advection Dominated Accretion Flows (ADAF) of
\citep[]{NY94}).

{\bf{\em 2. Winds:}} Analogous considerations of similarity apply
equally well to winds, given that, generally, their asymptotic
velocity, $v_{\infty}$, is proportional to the Keplerian velocity at
the wind launching radius, $r_l$, i.e. $v_{\infty} \propto
x_l^{-1/2}$, with $x_l = r_l/r_S$. So mass conservation in physical
and dimensionless units reads respectively
\begin{equation}
\dot M \sim n \, R^2 \, v_{\infty} ~~~~{\rm and }~~~~ \dot m_W \, M
\propto n(x) \, x^2 M^2 \, x_l^{-1/2}
\end{equation}
with $\dot m_W$ denoting in this case the mass outflow rate in units
of the Eddington rate $\dot M_E$. Therefore, the wind density and
local column density are given in dimensionless units
correspondingly by the expressions (dropping the subscript $W$ from
$\dot m_W$)
\begin{equation}
n(x) \propto  \frac{\dot m}{M}\frac{x_l^{1/2}}{x^2} ~~~~{\rm
and}~~~~ N_H(x) \propto \dot m \, \frac{x_l^{1/2}}{x} ~~.
\label{eq:column}
\end{equation}
with the object mass $M$ again dropping out of the expression for
the column density $N_H$. For 1D winds, e.g. stellar winds or winds
driven by the intense radiation pressure near a compact object,
$x_l$ is roughly constant and $N_H(x) \propto \dot m /x$, i.e. the
wind column decreases inversely proportionally to the distance.
However, in the more general case of 2D winds (e.g. those of BP82
and CL94) where the wind is launched from a wide range of radii in
an accretion disk with, generally, different mass flux rates at each
radius, $x_l \sim x, \, \dot m = \dot m(x)$ and $N_H(x) \propto \dot
m(x) x^{-1/2}$ and numerically $N_H(x) \simeq 10^{24} \dot m(x)
x^{-1/2} \, \cs$ (it should be understood that $n(x), \, N_H(x)$
have an additional angular dependence discussed in the previous
section). As noted in \cite{BB99} winds with position dependent
$\dot m$ are the means by which 2D accretion flows dispose of the
excess energy and angular momentum transferred {\em mechanically} by
the viscous stresses from the smaller to the larger radii, so that
finally  only a small fraction of the available mass accretes on the
compact object (BB99). For wind density of the form $n(x) \propto
x^{-s}$, the mass flow rate varies with $x$ like (BB99)
\begin{equation}
\dot m_W(x) = \dot m_{o} \, x^{-s +3/2}, \; s \le 3/2,
\label{eq:mdot}
\end{equation}
with $\dot m_{o}$ being the mass outflow from the smallest radius $x
\sim 1.5 - 3$; for $s=1, ~ \dot m \propto r^{1/2}$ and $N_H$
constant per decade of $r$.

{\bf{\em 3. Photoionization:}}  The wind ionization is determined by
the local ratio of photons to electrons, the ionization parameter
$\xi = L/n(r) r^2$, ($L$ is the source's ionizing luminosity --
different from the total luminosity, $n (r)$ the local density and
$r$ the distance from the ionizing source). This can also be
expressed in dimensionless units:
If $\eta \, ( \simeq 10\%)$ is the radiative efficiency of the
accretion process, then the luminosity $L$ can be written as $L
\propto \eta \, \dot m_a \, M$ ($\dot m_a$ is the accretion rate
that reaches the compact object to produce the luminoisty $L$), or
$L \propto \eta \, \dot m_a^2 \, M$ in the case of ADAF \citep{NY94}
[i.e for $\dot m_a \lsim \alpha^2$ with $\alpha$ the disk viscosity
parameter], yielding the for $\xi$ an expression also independent of
$M$, implying similarity in wind ionization, whether in AGN or XRB
(FKCB10)
%
\begin{eqnarray}
\xi(x) \simeq \frac{L}{n(r) r^2}  \simeq
\left\{ \begin{array}{ll} 
\displaystyle \frac{\eta \dot
m_a}{N_H(x)x} \simeq
 10^8 \displaystyle \frac{\eta}{f_W} \displaystyle \frac{1}{x^{-s+2}}  ~~~ &\mbox{ for $\dot m_a > \alpha^2$ (non-ADAF)} \\
\displaystyle\frac{\eta \dot
m_a^2}{N_H(x)x} \simeq
10^8 \displaystyle \frac{\eta }{f_W}\displaystyle \frac{ \dot m_a}{x^{-s+2}}  ~~~ &\mbox{ for $\dot m_a < \alpha^2$ (ADAF)} \\
\end{array}\right.
\label{eq:xi}
\end{eqnarray}
where $f_W = \dot m_{o}/\dot m_a \, (\sim 1)$ is the ratio of mass
flux in wind and accretion at the smallest radii and $s$ the index
of the density  dependence on $r$.  These relations are also
independent of $M$, implying a similarity in the ionization
structure of winds, whether in AGN of any type or GBHC. For a
spherical wind, ($s=2$) the ionization parameter is independent of
$x$ (asymptotically), while for $s=1$ it decreases linearly with
distance from the ionizing source.

{\bf{\em 4. The AMD:}} Writing Eq. (\ref{eq:xi}) as $N_H(x) \propto
\eta \, \dot m_a /
\xi(x) x$  one can form the expression for AMD (HBK07), namely
\begin{equation}
AMD = \frac{d N_H(x)}{d{\rm log} \xi(x)} \simeq \frac{\eta \, \dot
m_a}{\xi(x) x}
\end{equation}
The fact that AMD is largely independent of $\xi(x)$ (HBK07) implies
$\xi(x) \propto 1/x$ or $N_H(x) \propto log(x) $, $n(x) \propto 1/x$
and $\dot m \propto x^{1/2}$, i.e. the wind mass flux increases with
radius, as discussed in BB99. In this case, both the ionization
parameter and the wind density decrease like $1/r$ while the column
of the ions found at a given $\xi$ remains roughly constant, in
broad agreement with \citet{B09}. So, in disks of sufficiently large
radial extent, the wind launched at their larger radii will be of
sufficiently low ionization and temperature  to conform with the
properties of the AGN unification torus. Finally, one should bear in
mind that the above scaling laws do not include the
$\theta-$dependence of the density and the column density (see fig.
\ref{fig:fig1}). At a given (radial) distance $r$ from the source,
this leads to much higher ionization of the plasma flowing along the
symmetry axis and even smaller neutral absorption column than along
the higher column density, less ionized equatorial directions, in
agreement with the unification considerations  \citep{KK94}.

Under the above scaling laws, the ionization, velocity and AMD of
the winds discussed in this work are independent of the mass of the
accreting object. Therefore  these expressions should hold equally
well for AGN and GBHC or XRB in general, a fact in obvious
disagreement with observation, given their very different X-ray
absorber properties. This would indeed be the case if the ionizing
spectrum were also independent of the mass of the accreting object.
However, {\sl the similarity of wind ionization is finally broken by
the dependence of the ionizing (X-ray) flux on the mass of the
accreting object and leads to the different wind ionization
properties not only between XRB and AGN, but also amongst AGN:} Both
in AGN and XRB a major fraction of their bolometric luminosity is
emitted in quasithermal features referred to as the Big Blue Bump
(BBB) in AGN and the Multi Color Disk (MCD) in XRB. Both are thought
to represent emission of the accretion luminosity in black body form
by a geometrically thin optically thick disk of size a few \sw
radii. The characteristic temperature of this feature scales as
$M^{-1/4}$, being in the X-ray band for solar mass objects and in
the UV for AGN of $M \sim 10^8$ M$_{\odot}$. This dominance of the
ionizing luminosity (X-rays) as a fraction of the bolometric one in
XRB over AGN is, according to our model, the main cause for their
differences in their wind ionization properties as it will be
discussed in detail in the next section. Even within AGN,  the
ionizing flux is in fact not a constant fraction of the bolometric
luminosity but it depends on the absolute source luminosity. A
measure of their ionizing luminosity is given by the index
$\alpha_{OX}$, the logarithmic slope of the flux between 2500 $\aa$
and 2 keV. This quantity has a strong dependence on $L(2500 \aa)$
\citep{Steffen06,Brandt09}, a fact that bears relation to the
corresponding AGN absorber properties, as it will be discussed in
the next section.

\begin{figure}[t]
\begin{center}$
\begin{array}{cc}
\includegraphics[trim=0in 0in 0in
0in,keepaspectratio=false,width=2.2in,angle=-90,clip=false]
{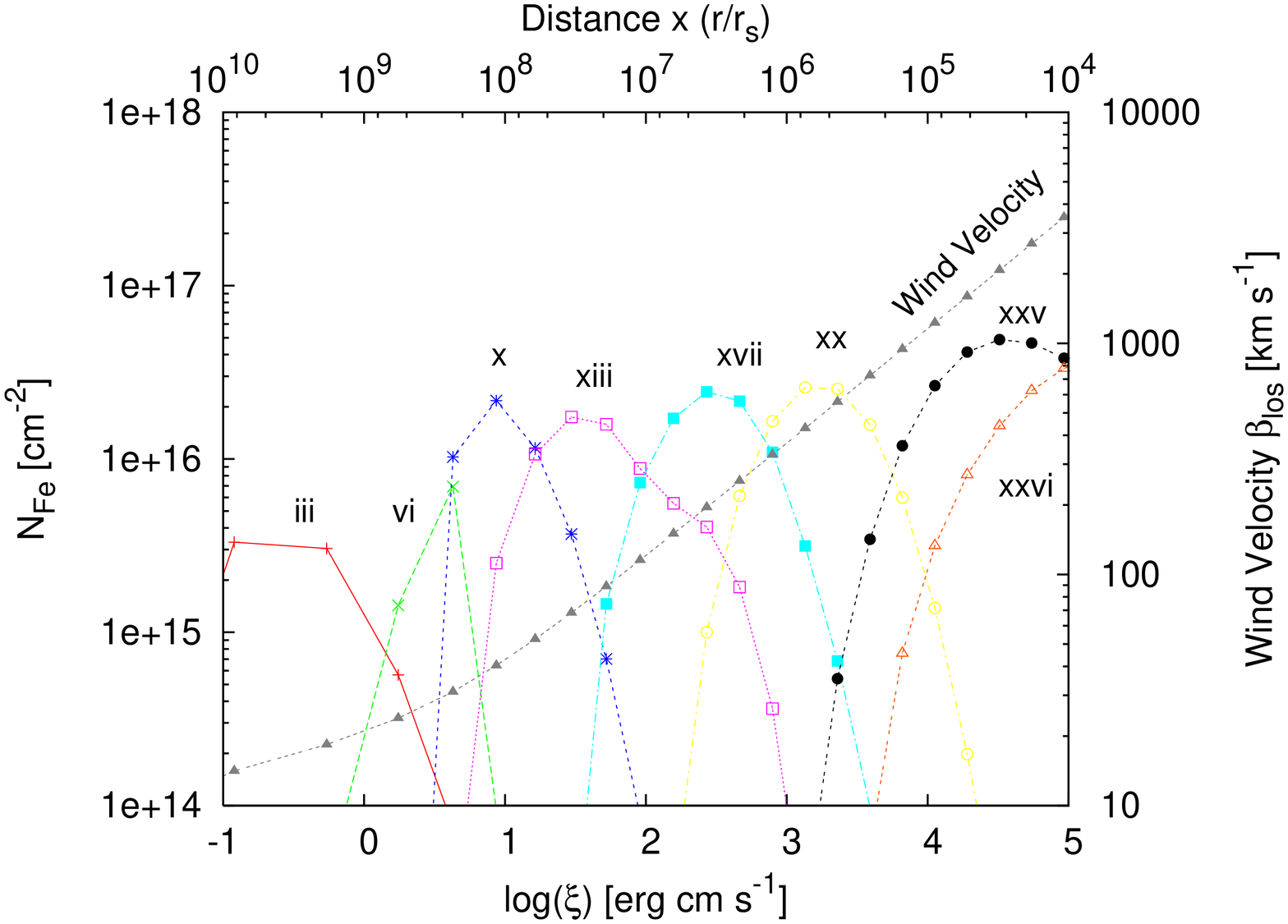} &
\includegraphics[trim=0in 0in 0in
0in,keepaspectratio=false,width=2.2in,angle=-90,clip=false]
{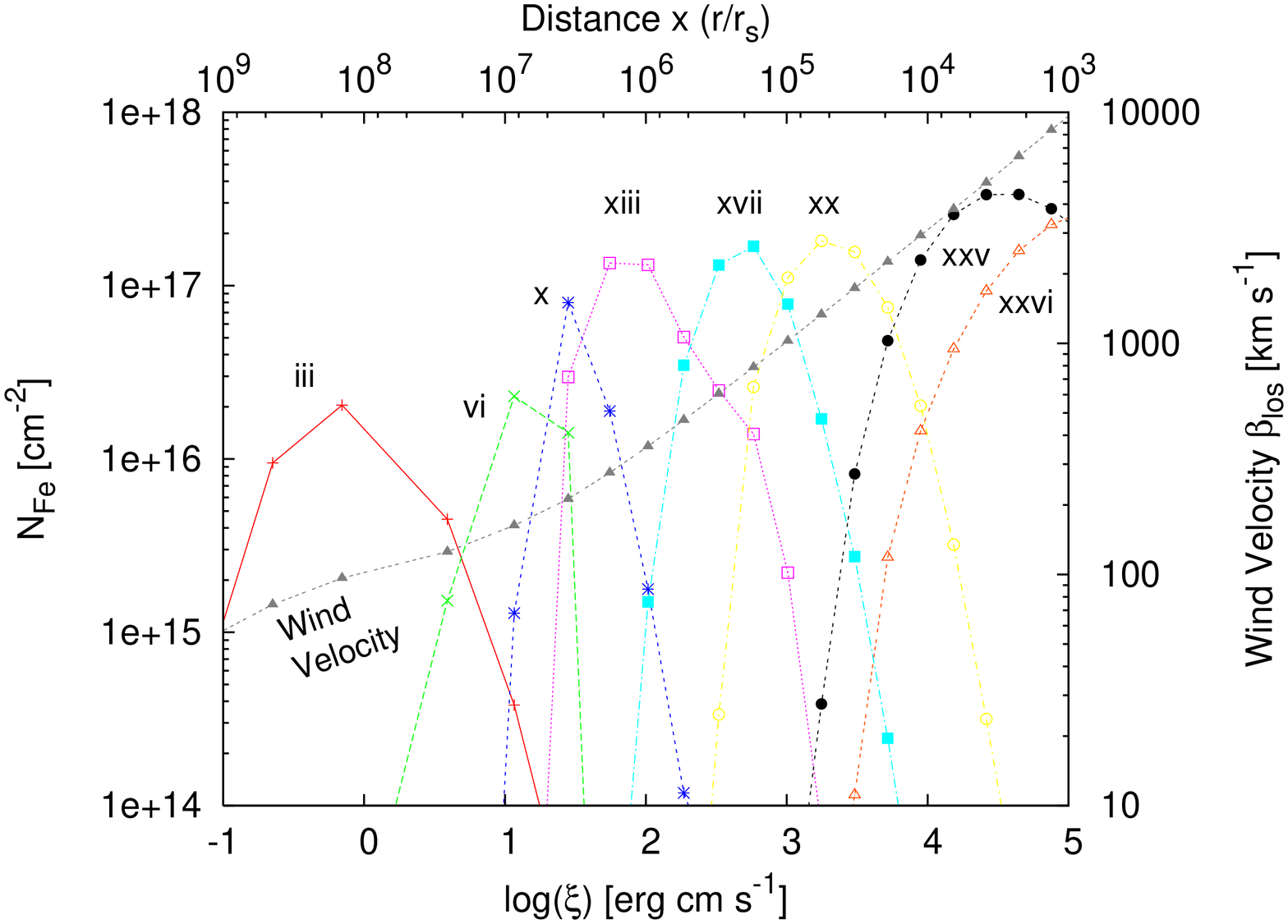}
\end{array}$
\end{center}
\caption{\small The ionization structure of iron appropriate for a Seyfert
galaxy with $\dot m_o \simeq 0.1$ and an ionizing photon spectrum with
$dF_{\nu}/d \nu \propto \nu^{-\Gamma}, ~ \Gamma = 1.9$, as seen at two
different inclination angles $\theta = 30^{\circ}$ (left) and
$\theta = 60^{\circ}$ (right). The ionization decreases with
increasing distance $r$ along a given LoS and with increasing
$\theta$ for a given value of $r$. The column of each ion is given
in the left ordinate, while the velocity at each position along the LoS
in the right one. The LoS velocity at each $\xi$ (lower abscissa)
and the corresponding radius (upper abscissa) is given by the black
triangles.} \label{fig:fig2}
\end{figure}

\section{The Ionization Structure of MHD Winds}

With the wind structure set by the solution of the Grad-Safranov
equation, one can now proceed to study its ionization structure. The
details are given in \citet{FKCB} so we simply restrict ourselves to
a broad description of this procedure. Given the (approximately)
$1/r$ density profile, we have split the radius along a given LoS
from $r_o$ to $ 10^6-10^8 \, r_o$ in approximately 40 segments,
spaced equally in log$r$, so that each has the same column density.
Then we assume a value for $\dot m_o$ which provides the
normalization of the wind density through Eq. (\ref{eq:dnorm}) (we
generally assume $f_W =1$) and a luminosity $L$ related to $\dot
m_o$ either as $L = 1.3 \cdot 10^{38} \, \dot m_o M$ erg/s or $L =
1.3 \cdot 10^{38} \,\dot m_o^2 M$ erg/s (for $\dot m_o < \alpha^2$,
according to the ADAF prescription) with a specific spectrum; then
we invoke the photoionization code \verb"XSTAR" to compute the
ionization, opacity and emissivity of the first zone. Using the
derived opacities and emissivity we compute the spectrum exiting
this zone, which is used as the input for computing the ionization
opacity and emissivity of the next zone. The procedure is repeated
to compute the ionization of the wind along a given line of sight
and then for different values of the polar angle $\theta$, to
produce the ionization structure of the wind over the entire
poloidal plane.

Fig. 3 of \citet{FKCB} depicts the density and ionization structure
of such winds out to $10^8 \, r_o$ in linear coordinates. One can
see that the wind ionization depends both on the distance from the
source $r$ and on  the angle $\theta$, being highly ionized at small
values ($\xi \gsim 10^3$ near $\theta = 0$) even at these large
distances, while it decreases to $\xi < 1$ at $\theta \sim
80^{\circ}$.
It is also clear that for a given $\theta$, a harder X-ray spectrum
will ionize the wind out to larger distances. Since the wind
velocity along a given LoS depends mainly on the radius of launching
the specific parcel of gas (but also the inclination angle
$\theta$), absorption features associated with a given ion should
appear at smaller velocities the harder the ionizing spectra. This
issue is in fact a little more complicated because we only observe
the wind velocity along our line of sight which depends also on the
local shape of the poloidal magnetic field lines.

\subsection{Ionization of XRB Winds}

The high X-ray content of the spectra of XRB, implies that they
should be, in general, much more ionized than those of AGN.
Nonetheless, X-ray absorption features have also been observed in
XRBs, as indeed seems to be the case. So P-Cygni Fe-K features were
first  detected in Cir X-1, which is (presumably) an accreting
neutron star in a 16.6 day binary orbit around a main sequence
companion \citep{BS00,SB02}. The high ionization of the inner
sections of these outflows, render them essentially devoid of any
absorption features. Therefore, the highest ionization species
\fexxvi, \fexxv~should occur at radii such that $\xi \lsim 10^4$,
which, as the authors suggest sets them at distances $r \simeq 10^5$
km and corresponding  velocities of only $v \simeq 1,000$ km/s.

There are several other XRB in  which similar features have been
detected, the best known being GRO 1655-40 \citep{Miller06,Miller08}
and GRS 1915+105 \citep{NRL} both of which show \fexxvi~and \fexxv~
absorption features at velocities similar to those of Cir X-1.
GRO 1655-40 presents a multitude of absorption features similar
to those of AGN, however, because of the higher S/N ratio their
velocities and EW are better determined and thus more contraining
to the outflow models. Following detailed modeling it was concluded
\citep{Miller06,Miller08} that the wind of GRO 1655-40 is not consistent
with being driven
by either line radiation pressure or  thermally by the X-ray heating
of the outer regions of the disk. The authors then concluded that
the only reasonable means of driving the  outflow in this object
is magnetic, via an MHD wind similar to those outlined in this
work.

GRS 1915+105 affords a smilar analysis \citep[][and references
therein]{NRL}. Detailed photoinization modeling of this source
provided both the column and velocity associated with specific
transitions. Assuming that the line absorbing medium has  thickness
$\Delta r \simeq r$, and that the observed velocities are
effectively Keplerian, i.e. $v/c \simeq (r/r_S)^{-1/2}$, one can
obtain simultaneous estimates of both $r$ and $v$; these along with
the measured absorption column  $N_H \simeq n r$ yield an estimate
of the  mass flux of the wind, which it was found that it could  be
as much as several times larger than the mass accretion rate needed
to power the observed X-ray emission of this source. In fact, one is
led to a similar conclusion concerning the mass flux in the wind of
Cir X-1 by using the parameters provided in \citet{SB02}, indicating
that a wind mass flux increasing with $r$ may not be  an uncommon
feature in accretion powered sources.

Finally, it is worth noting that the disks of all three of the above
sources are apparently at high inclinations, i.e. directions that
favor observations of such features (less ionization, higher column
densities). For lower inclination systems, the winds are likely far
more highly ionized, yielding no apparent absorption features.

\subsection{Ionization Structure of Seyfert Winds}

The  procedure outlined in \S 4 describing the detailed calculation
of the ionization of MHD winds was applied to winds associated with
Seyfert galaxies \citep{FKCB}. As noted above, given the scale
invariance of the wind column and ionization parameter, the main
difference between the ionization winds off the disks of different
classes of accretion powered sources (e.g. XRB, Seyferts, BAL QSOs)
is the spectrum of the ionizing radiation. Since for Seyfert
galaxies the X-ray and UV-optical luminosities are approximately the
same, we have approximated the ionizing spectra with a single power
law from $1 - 1,000$ Rydberg of photon index $\Gamma = 1.9$ and for
the accretion rate we used the value of $\dot m_o = 0.1$.

The 2D  density and poloidal plane ionization structure of a wind
for the above parameters is given in Fig. 3 of \citet{FKCB}. Here,
we present in Figure \ref{fig:fig2}  the ionic species distribution
of just one element of this wind, namely iron, as a function of the
distance from the continuum source (upper abscissa) and the
corresponding value of $\xi$ (lower abscissa), for two different
values of the observer inclination angle $\theta= 30^{\circ}, \;
60^{\circ}$ (left and right panels respectively). The column of each
ionic species is given in the left ordinate, while the corresponding
velocity along the LoS is given in the right ordinate and
represented by the black triangles.
%

We can see that increasing $\theta$ increases the column of a
specific ionic species. Also, because the radius at which it now
forms is smaller, the velocity at the peak abundance of this ion is
correspondingly higher. Finally, one can obtain the AMD of this flow
by adding the columns of all ions of Fe at a given value of $\xi$.
As expected, and shown explicitly  in  \citep{FKCB}, this is
constant (i.e. independent of the value of $\xi$ or the value of
$r/r_S$), reflecting the column per decade of radius of the
underlying wind.

The ionization properties of Seyfert galaxies are consistent with
the models presented in this figure. The conclusions of \citet{B09},
based on the AMDs of the five AGN he has analyzed are broadly
consistent with ours as presented in this section. Some of them
require slightly steeper density profiles ($s \simeq 1.25$) whose
ionization structure and velocity properties should be similar to
those presented in this section.

\begin{figure}[h]
\begin{center}$
\begin{array}{cc}
\includegraphics[trim=0in 0in 0in
0in,keepaspectratio=false,width=2.7in,angle=-0,clip=false]
{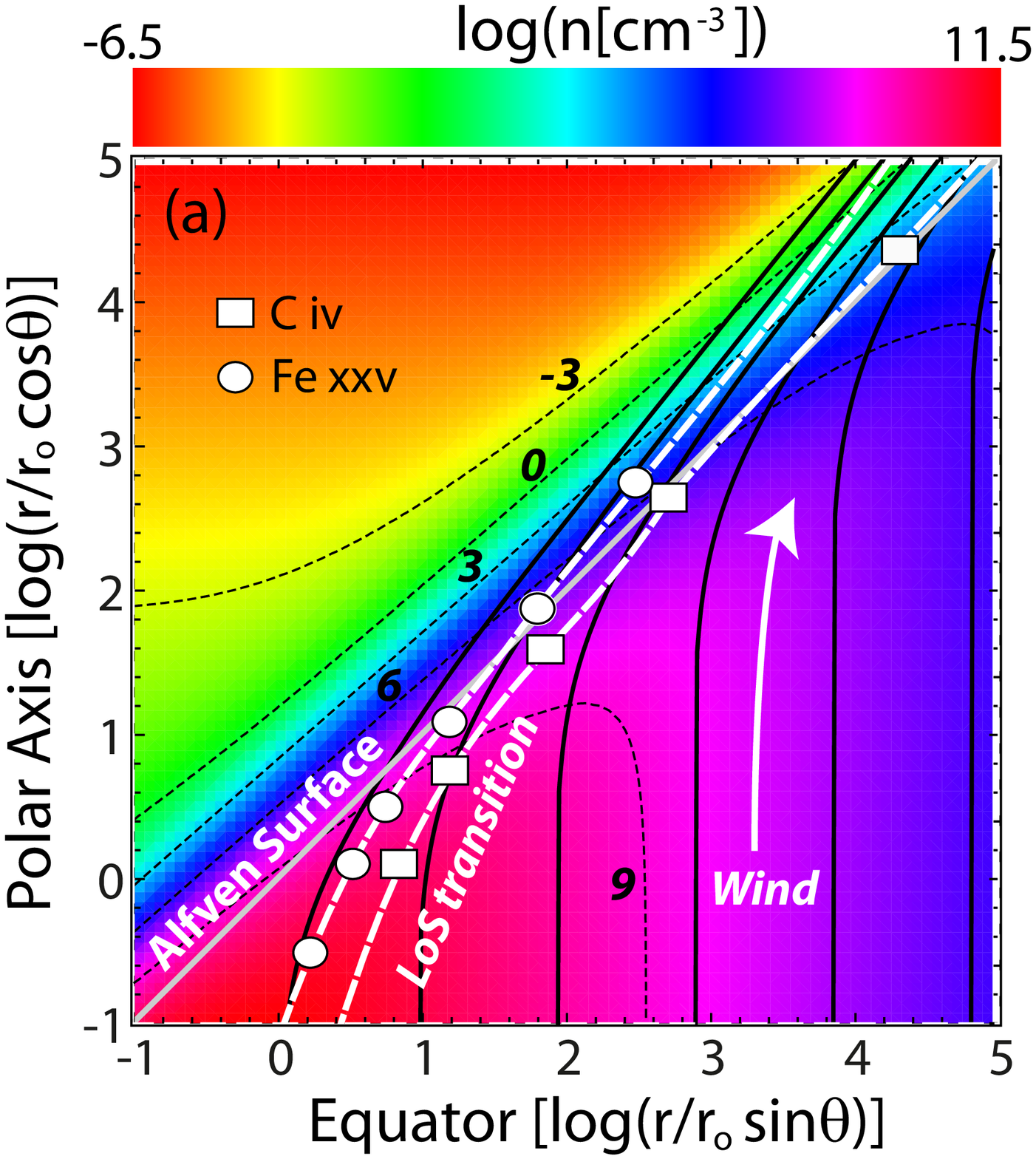} &
\includegraphics[trim=0in 0in 0in
0in,keepaspectratio=false,width=3.3in,angle=-0,clip=false]
{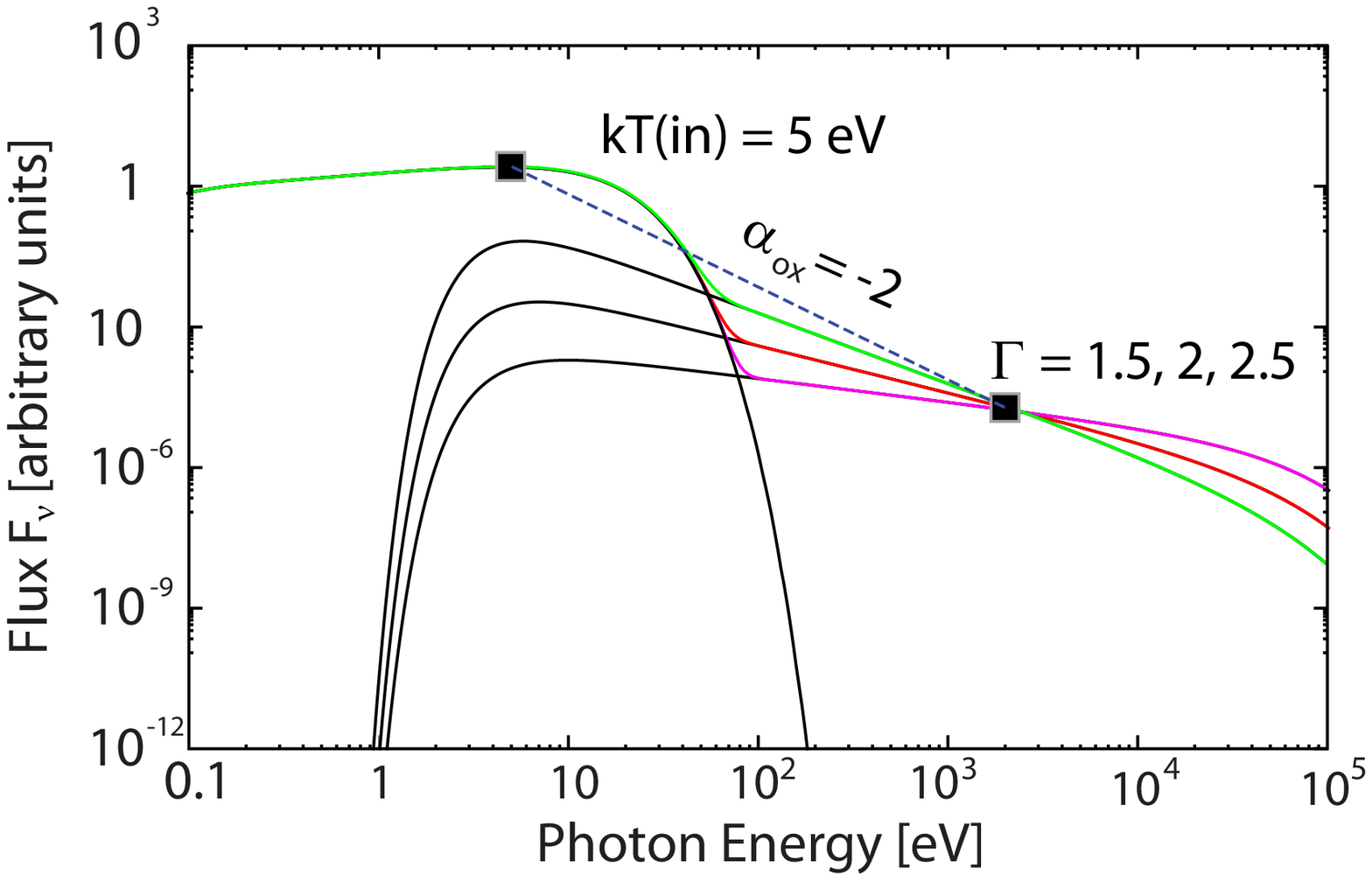}
\end{array}$
\end{center}
\caption{\small (a) 2D poloidal plane density structure of an MHD wind of
$\dot m = 0.1$ and $M=10^9 \Msun$, ionized by the spectrum shown in
(b). The density, $\log (n[\textrm{cm}^{-3}])$, is shown in color
with the dotted lines being the iso-density contours, with the log
of the corresponding density indicated by the associated numbers.
The $r, z$ coordinates are in logarithmic scale, so lines of
increasing $\theta$ are lines at 45$^{\circ}$ inclination of
increasing abscissa intercepts. Also shown are the magnetic field
lines ({\it black solid}) and the loci ({\it white dashed lines}) of
the positions of formation of \fexxv~(circles) and \civ~(squares).
(b) The form of the SED used to produce the wind ionization of (a)
consisting of (i) a thermal MCD (as BBB) spectrum of innermost
temperature $k T_{\rm in}=5$ eV and (ii) a power-law continuum of
photon index $\Gamma$. The X-ray flux is normalized relative to that
of BBB by $\alpha_{\rm ox}$. [{\it See the electronic edition of the
Journal for a color version of this figure.}]} \label{fig:fig3}
\end{figure}

\subsection{Ionization Structure of BAL QSO Winds}

If the tenets of the proposed model are correct, it should be able
to reproduce the velocity and ionization structure of the absorption
features of accretion powered objects other than Seyfert galaxies
through judicious choices of the values of the available parameters.
Should such an enterprise be successful, it could delineate the
regions of parameter space occupied by individual AGN classes and
hence provide their systematization on the basis of an
underlying physical model.

Such an attempt was made by  \citet{FKCB1}, who employed the
specific MHD winds to model the absorption features in the spectra
of BAL QSOs, in particular that of APM~08279+0255
\citep{Chartas02,Chartas03}. This AGN class was chosen because the
properties of  absorption features of its members  are  most
different from those of Seyferts, since they exhibit \fexxv~at
velocities $v\simeq 0.5 c$ and \civ~at $v\simeq 0.1c$. Because these
objects were selected on the basis of their broad UV absorption
features, any such model should be able to provide an account of the
{\em combined} properties of both their X-ray and UV features.

The property that most clearly separates Seyferts and BAL QSOs is
the relative weakness of their X-ray to their UV luminosity. This is
reflected in the corresponding value of their $\alpha_{OX}$
parameter, which in BAL QSOs has values $\alpha_{OX} \simeq -2$
compared to $-1$ to $-1.2$ for Seyferts. To model the ionization of
BAL QSO winds, \citet{FKCB1} have chosen as ionizing spectrum that
shown in Fig. \ref{fig:fig3}b. This consists of a BBB component of
maximum temperature 5 eV along with a power law component of photon
index $\Gamma$, normalized at $E=2$ keV relative to the BBB so that
it results in $\alpha_{OX} = -2$, consistent with that of
APM~08279+0255. The decrease in the value of $\alpha_{OX}$ is
significant, because it implies that the ratio of ionizing photon
density (given by the X-rays) to the matter density (represented by
the luminosity of the BBB that peaks in the UV) is quite small,
yielding a value for the ionization parameter at $r\simeq r_o$ much
smaller than that of Seyfert galaxies. Furthermore, the cooling
effects of the prominent UV emission further contributes to lowering
the gas temperature at the smallest radii and hence its ionization.
It is then possible that some of the heavy elements (e.g.  Fe) will
not be fully ionized even at wind radii $r \simeq r_o$, leading to a
high outflow velocity for \fexxv,~\fexxvi.

The presence of non-fully ionized Fe at radii $ r \gsim r_o$, along
with the very steep (unobservable) EUV source spectrum, depletes
severely the photons at energies $E \gsim 50$ eV needed to produce
the ion $C^{+3}$ of carbon. Since this ion forms at a specific value
of $\xi$, this suppression of ionizing photons is made up by a
decrease in the radius $r$ at which it forms (and an increase in the
local wind velocity), leading to \civ~absorption features at
velocities much higher than those of Seyfert galaxies. Figure
\ref{fig:fig3}a shows in logarithmic scale the density distribution
of a wind of $\dot m_o =0.1, ~{\rm and}~ {\rm M} =10^9 \, {\rm
M}_{\odot}$. Dashed lines are the isodensity contours with the
logarithm of the corresponding density denoted by the assigned
number, while the thick black lines are the lines of the poloidal
magnetic field. The wind flow in the poloidal plane is  along the
magnetic field lines, while the observers' LoS (radial coordinates)
correspond to lines of 45 degree inclination; an increasing
inclination angle $\theta$ corresponds to increasing abscissa
intercepts for these lines. The white circles and squares along with
their connecting curves indicate correspondingly the positions of
formation of the \fexxv~and \civ~transitions for observers at
different values of $\theta$. {The ionization structure of the wind
can  thus be translated into the spatial localization of specific
ionic species. With a judicious choice of $\theta\; (\simeq
50^{\circ})$ one can obtain $r($\fexxv$) \sim 5-40\; r_S$ and
$r($\civ$) \sim 300-900\; r_S$ (assuming a single LoS)} to produce
absorption velocities for these transitions consistent those of the
BAL QSO APM~08279+0255.

\begin{figure}[h]
\begin{center}$
\begin{array}{cc}
\includegraphics[trim=0in 0in 0in
0in,keepaspectratio=false,width=3.1in,angle=-0]
{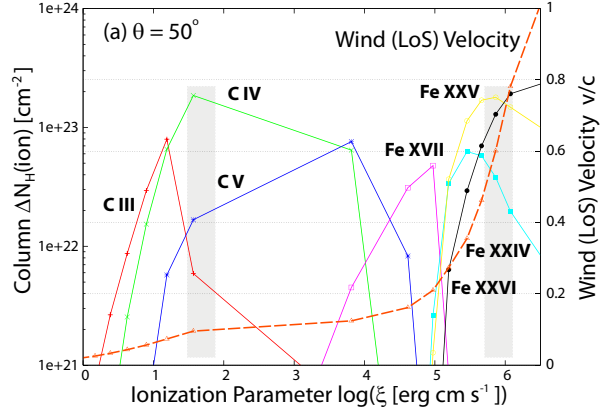} &
\end{array}$
\end{center}
\caption{\small Simulated distribution of local column densities $\Delta
N_H$ (on the left ordinate) and the outflow LoS velocity $v/c$ (on
the right ordinate) as a function of ionization parameter $\xi$ for
(a) C and (b) Fe with $\theta=50\deg$. Vertical lines denote the
ionization parameter where the local emergent column is dominated
primarily by \civ ~in (a) and \fexxv ~in (b), respectively, also
showing the corresponding outflow velocity $v/c$. [{\it See the
electronic edition of the Journal for a color version of this
figure.}]} \label{fig:fig4}
\end{figure}

In Fig. \ref{fig:fig4} we show the hydrogen equivalent columns of
several ions of iron and carbon as a function of the ionization
parameter $\xi$ for a LoS at an  angle $\theta=50^{\circ}$ and for
the ionizing spectrum given in Fig. \ref{fig:fig3}b. This is a
figure equivalent to those for Seyferts shown in Fig. 2 but
appropriate for a BAL QSO. The figure shows also the
corresponding wind velocity along this LoS (red dashed line), while
the shaded regions indicate the positions of maximum column for
\fexxv~and \civ, which are indicative of the observed velocity of
the corresponding ion.

\subsection{The Characteristic Curves of Ionized MHD Winds}

The previous two subsections have shown that, while the underlying
velocity structure of MHD winds may be very similar and their
overall column depending only on the dimensionless mass flux $\dot
m_o$, their ionization structure and the location of specific ions
in these winds can be quite different depending on their ionizing
spectra. Under these (admittedly oversimplified) assumptions one can
present the kinematic properties of specific ionic species of these
winds into two  diagrams that encapsulate their essential features
(namely velocity and column) as a function of the ionizing radiation
spectral properties and the observers' LoS.

\begin{figure}[t]
\begin{center}$
\begin{array}{cc}
\includegraphics[trim=0in 0in 0in
0in,keepaspectratio=false,width=2.9in,angle=-0] {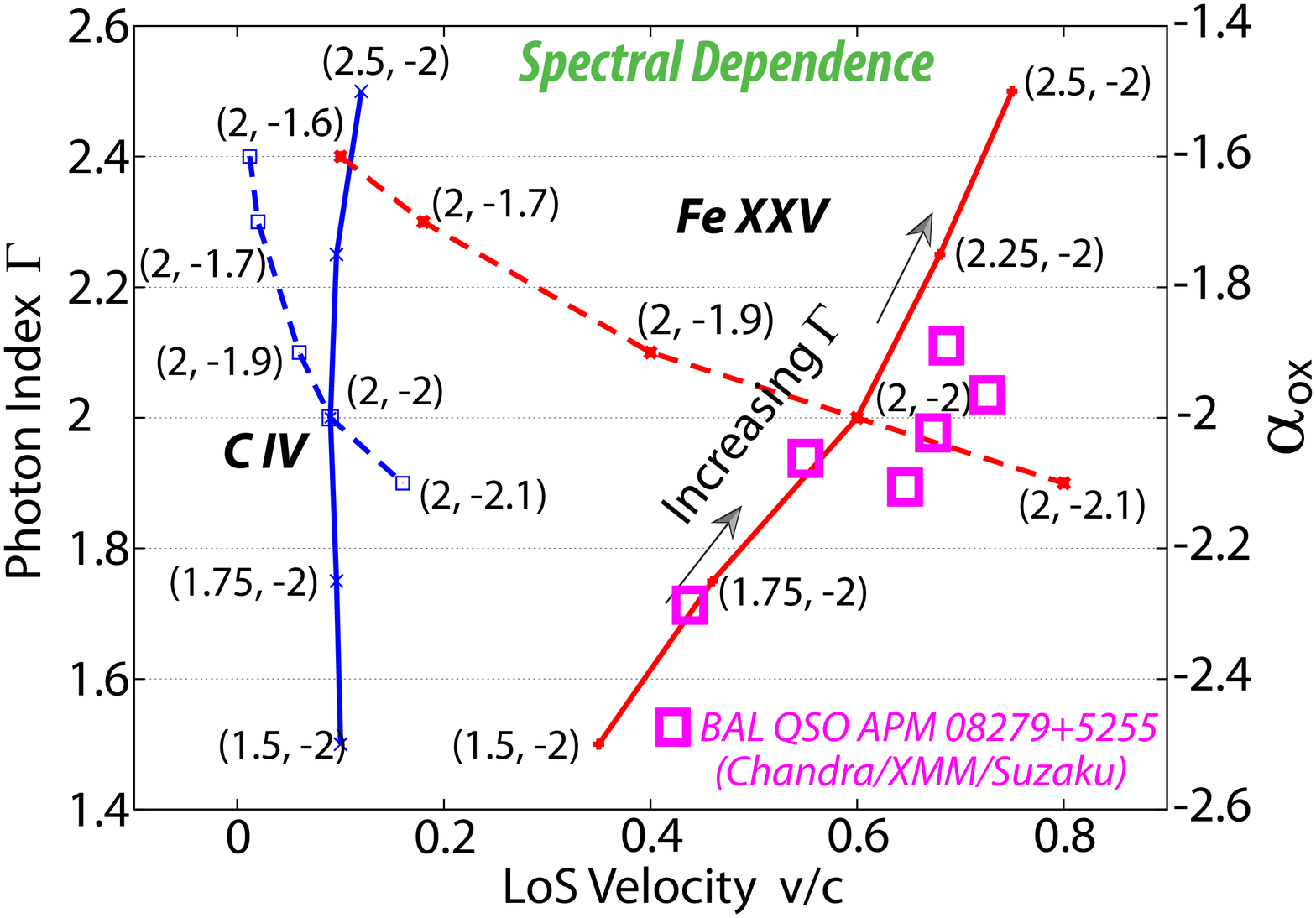} &
\includegraphics[keepaspectratio=false,width=2.9in,angle=-0]
{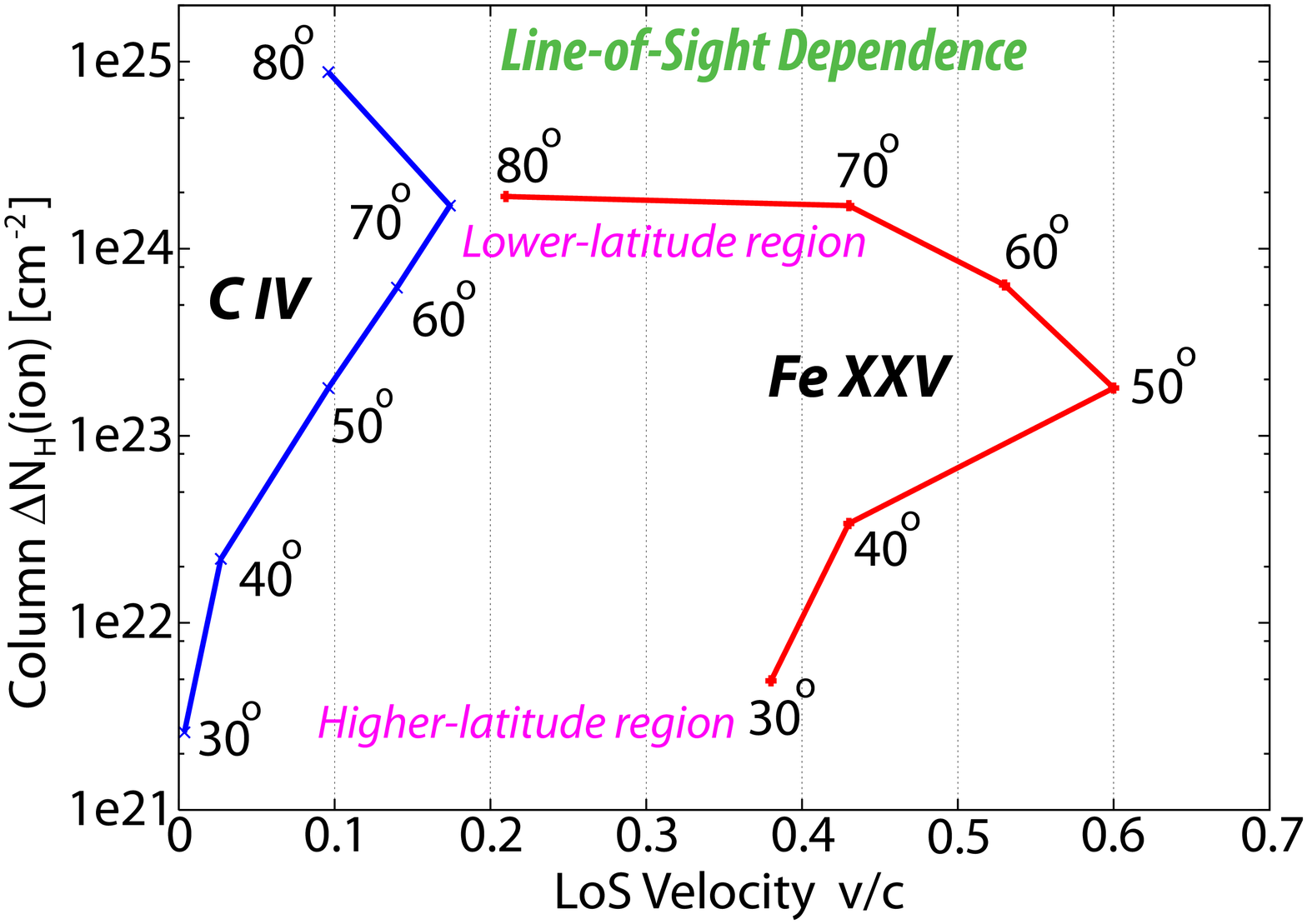}
\end{array}$
\end{center}
\caption{\small (a) The dependence of the model wind LoS velocity (for
$\theta = 50^{\circ}$) on $\Gamma$ ({\it solid lines; left
ordinate}) and $\alpha_{ OX}$ ({\it dashed lines; right
ordinate}). The ordered pairs at each point are the values of
($\Gamma,\alpha_{OX}$) for \civ~(blue lines) and \fexxv~(red
lines). The red squares indicate the change in the Fe-K absorption
velocity variations with $\Gamma$ for APM 08259+5255. (b) The model
MHD wind  correlation between the LoS velocity of a given transition
(blue for \civ, red for \fexxv) and the corresponding hydrogen
equivalent column $\Delta N_H$(ion) with the observer inclination
angle $\theta$ as a parameter along these curves (indicated by the
numbers in degrees). [{\it See the electronic edition of the Journal
for a color version of this figure.}] } \label{fig:fig5}
\end{figure}

Such diagrams for the transitions \civ($\lambda\lambda
1548,1551$\AA) and \fexxv~are shown in Figs. \ref{fig:fig5}a,b. The
\civ~characteristic curves are shown in blue while those of
\fexxv~in red. In Fig. \ref{fig:fig5}a, appropriate for an observer
at $\theta = 50^{\circ}$, we show: (i) The correlations between the
\civ,~\fexxv~velocities and the X-ray photon index $\Gamma$ for
$\alpha_{OX}=-2$ (solid lines, left ordinate). (ii) The correlation
between the \civ,~\fexxv~velocities and the source $\alpha_{OX}$ for
an X-ray photon slope $\Gamma = 2$ (dashed lines, right ordinate).
The pairs in parentheses indicate the values of ($\Gamma,
\alpha_{OX}$) at each point.

We observe that the velocities of both \civ~and \fexxv~depend more
strongly on $\alpha_{OX}$ than on $\Gamma$. However, for
$\alpha_{OX}=-2$, as is the case with APM~08279+0255, the
\fexxv~velocity changes significantly with $\Gamma$ and in  a way
that is actually  in agreement with the observations of
\citet{Chartas09}, given by the squares along the solid red curve.
The \civ~velocity appears to be much less sensitive to $\Gamma$,
however it is quite sensitive to the value of $\alpha_{OX}$, ranging
between $v \simeq 0.15c ~ {\rm and} ~0.01c$ as $\alpha_{OX}$ varies
from $-2.1$ to $-1.6$, nicely covering the observed range of the
velocity of this transition between BAL QSO and Seyferts.

In Fig. \ref{fig:fig5}b we show the correlation between the LoS
velocity of the specific ions (i.e. \civ, ~\fexxv) for a wind
ionized by a source  with $\alpha_{OX}=2, ~\Gamma =2$ (values
appropriate for a BAL QSO),  and the corresponding hydrogen
equivalent columns, $N_H$, with the LoS inclination angle as a
parameter along these curves. For these specific spectral
parameters, we see that both velocities increase with $\theta$ up to
a critical value different for each transition and then level-off as
the corresponding ions are produced in parts of the wind that it is
still accelerating. It is also worth noting that even for the
specific spectral distribution, appropriate for BAL QSOs, the
velocity of \civ~is quite small for sufficiently small observer
inclination angle, while at the same time the corresponding column
may be too small for the detection of such a feature. Diagrams such
as that of Fig. \ref{fig:fig5}b for different ionizing spectra and
wind parameters will be useful for the determination of the wind
parameter space ($N_H, v$) at which specific features become
observable. We expect to provide several such diagrams for the most
important such transitions in future works.

\section{Wind Emission Properties}

The models and computational setup outlined in the previous were
motivated by and dealt mainly with the absorption feature properties
of accretion powered sources, both galactic and extragalactic.
However, besides the absorber properties, the reprocessing of
radiation onto these winds provides also emission characteristics,
both in continuum and in lines. These will be given a first brief
treatment in this section, which serves mainly as the road map of
future work. So in section 5.1 we discuss the effects of dust
reprocessing into continuum while in 5.2 reprocessing into lines and
specifically that of H$\alpha$.

\subsection{Torus Structure and Infrared Emission}

The issue of AGN torii and their importance in unification schemes
was discussed in the introduction. Also outlined  there was the
conundrum of the disparity between their (statistically inferred)
height-to-radius ratios ($h/R \simeq 1$) and the values implied by
the ratio of their thermal to Keplerian velocities ($h/R \simeq
v_{\rm th}/v_{\rm K} \simeq 10^{-3}$).
The proposal that resolves this conundrum in a straightforward
fashion is that of \citet{KK94} who suggested that the  torii  are
not structures in hydrostatic equilibrium, but dynamical objects, in
particular MHD winds driven-off the outer environs of the AGN disks.
Furthermore, in fitting the observed IR emission as disk UV
radiation reprocessed in these winds, they concluded that the
corresponding density of the reprocessing gas had to be proportional
to $1/r$ as in the winds discussed in \S 3.

\begin{figure}[t]
\begin{center}$
\begin{array}{cc}
\includegraphics[trim=0in 0in 0in
0in,keepaspectratio=false,width=3.1in,angle=-0]
{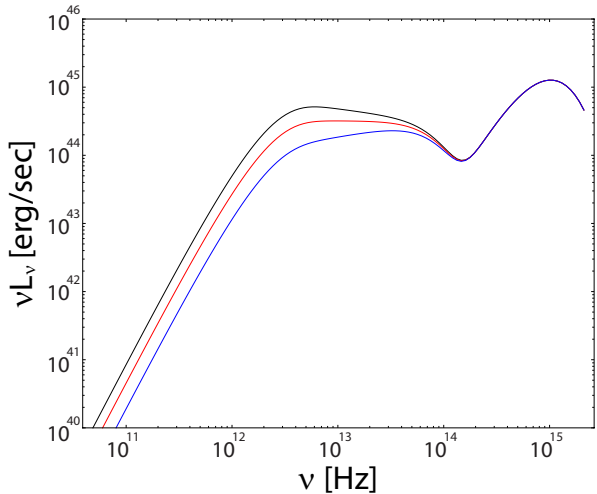} &
\end{array}$
\end{center}
\caption{Model spectra resulting from reprocessing the BBB AGN
continuum (with peak at $\nu \simeq 10^{15}$ Hz) by winds with
three different density profiles $s=0.9$ (black), $s=1.0$ (red),
$s=1.1$ (blue), that extend roughly two orders of magnitude
beyond the dust sublimation radius. [{\it See the
electronic edition of the Journal for a color version of this
figure.}]} \label{fig:fig7}
\end{figure}

Similarly, \citet{RR95} has shown that reprocessing the AGN
continuum luminosity by dust can provide good fits to their (nearly
flat in $\nu F_{\nu}$) IR spectra, provided that the dust density
has a profile similar to that used herein, namely proportional to
$1/r$. There is no mystery here: Dust distribution with the specific
density profile reprocesses the same amount of continuum flux per
decade of radius and re-emits it in (roughly) black body form; since
the dust temperature in regions of increasing radius $r$ falls-off
like $T \propto r^{-1/2}$ the peak emission will also shift to lower
frequencies while emitting the same luminosity; the resulting
spectrum would then be ``flat'' in $\nu F_{\nu}$ units, consistent
with the {\em Spitzer} observations of a survey of QSO spectra
\citep{Netzer07}.

The precise spectrum determination depends, of course, on the
details of dust formation and properties and also on the details of
the reprocessed spectrum and radiative transfer, as the emission is
not necessarily black body and as the outer layers of the MHD wind
reprocess not only the continuum luminosity but also the IR emission
of the inner ``torus'' regions. One of the more important aspects of
the IR and far-IR emission is the frequency at which the spectrum
turns over from ``flat'' to the Rayleigh-Jeans form ($\nu F_{\nu}
\propto \nu^3$), as this would provide an estimate of the outer
torus radius. Combined  space ({\em Herschel}) and ground ({\em
ALMA}) observations will be instrumental in this respect. One should
simply note here that the radial dust temperature dependence, $T
\propto r^{-1/2}$, implies that the number of decades in radius over
which dust reprocessing takes place is twice the number of decades
over which the AGN $\nu F_{\nu}$ spectrum is ``flat''. In this
respect, there should be a synergy between AGN X-ray spectroscopy
and far-IR observations: The regions at which the far-IR emission
turns over to the Rayleigh-Jeans form (the ``egde" of the torus) are
comparable to those at which the lowest $\xi$ ions should be
located; the ``edge" of the torus should then manifest then also in
the X-ray spectra as a steep drop in the X-ray column with
decreasing $\xi$.
Finally, one should note that IR spectra similar to those of
Seyfert-2 galaxies are not expected in XRB, despite their similarity
of their winds; this is because the distance at which the dust
sublimation temperature ($T \sim 1000$ K) is reached in the wind of
an XRB of $L = 10^{37}$ erg s$^{-1}$, is larger than $10^{12}$ cm,
the typical XRB binary separation distance.

Fig. \ref{fig:fig7} presents an oversimplified example of such a
spectrum. In there we present the reprocessing of the BBB spectrum
of an AGN (the component that peaks near $\nu \sim 10^{15}$ Hz),  by
dust which extends  from the radius of dust sublimation to the
``edge" of the torus, a region that spans roughly two decades in
radius. The dust density in this region has a form $n(r) \propto
r^{-s}$ with $s = 0.9, 1, 1.1$ (black, red and blue curves
respectively) to indicate the effect that different density profiles
on the far-IR AGN spectra. In this specific case the AGN is viewed
``face-on'' and the reprocessing is effectively limited to $\theta
\gsim 50^{\circ}$. The details of the X-ray obscuration and its
relation to IR emission, bearing relation to the angular
distribution of our solutions, can be improved with combined X-ray
-- IR surveys such as that of \citep{RR-N09}.

\subsection{H$\alpha$ Emission of MHD Winds}

The issue of line emission in AGN is a subject far too broad to be
included in any detail in the present work. Nonetheless, to our
surprise, we found that the basic model we have been discussing so
far is consistent with well known features and correlations
concerning the {\em emission line} properties of AGN and XRB. So we
do make a brief foray into this subject to touch upon two issues on
which our model seems to provide a novel insight.

One of the fundamental characteristics of the winds outlined in
section 3 are that their velocity fields are mainly Keplerian at low
latitudes ($\theta \gsim 70^{\circ}$) while becoming (almost) radial
at larger ones \citep{CL94,FKCB}. This particular property allows
for the possibility of exhibiting emission line features associated
both with disks and winds. In fact, there are observations of a
``double horn'' H$\alpha$ line profile in certain AGN \citep{EH94},
the tell-tale sign of disk emission. These have been modeled as
emission excited by the reprocessing of the AGN X-ray continuum on a
ring-like structure, whose radii and illumination were adjusted to
provide  agreement with observations.  The MHD wind models in their
most simplistic incarnation, namely that described \S 3, allow for
such profiles for the H$\alpha$ transition, precisely because of the
disk-like geometry of the flow at low latitudes. Furthermore, the
extent of the wind in the $\theta-$direction allows also the {\em
calculation} of the EW of this transition, which is found to be
consistent with observations (Fukumura \& Kazanas in preparation).

The calculations outlined in \S 4 provide the photonionization of
the entire wind, from its smallest ($x \gsim 1$) to its largest ($x
\sim 10^5 - 10^7$) scales, set by the size of the accretion disk
that powers the source. With the wind ionization given, one can also
produce scalings and correlations involving properties of the
corresponding emission of lines of the photoionized plasma. While
such an in-depth study is beyond the scope of the present paper,
some straightforward scalings can be produced even at this early
stage. With this in mind, we compute below  the scaling of one of
the prominent transitions AGN, namely H$\alpha$ with the source
luminosity, given that such a relation has already been well
documented \citep{Oster89} (we refrain focusing on the Ly$\alpha$
and \civ~transitions since, being resonant ones, demand a more
careful treatment within the specific model).

Quite generally, the luminosity of the H$\alpha$ transition produced
in the MHD winds discussed in the previous sections can be written
as
\begin{equation}
L_{{\rm
H}\alpha} \simeq \alpha(T)\, n^2 V \, \epsilon_{{\rm H}\alpha}
\end{equation}
where $\alpha (T)$ is the recombination coefficient, $n^2 V$ the
wind emission measure and $\epsilon_{{\rm H}\alpha} \simeq 2$ eV, is
the H$\alpha$ energy. As noted in Eq. (\ref{eq:dens}), the wind
density is given by the expression $n (r, \theta) \simeq n_o \tilde
n (\theta) x^{-1} \dot m / M$, where $\tilde n (\theta) \simeq
e^{(\theta - \pi/2)/0.22}$. The normalization $n_o$ is set so that
$n_o r_o \sigma_T \simeq 1$ for $\dot m \simeq 1$ in spherical
geometry. For the same value of the total mass flux, since the flow
in the MHD winds considered herein is concentrated near the
equatorial plane, the density normalization must be higher by a
factor $A \simeq 1/ \int_0^{\pi/2} \tilde n (\theta) \, {\rm sin}
\theta \, d \theta \simeq 1/0.21$ with the density profile now
reading $n (r, \theta) \simeq n_o \, A \,\tilde n (\theta)  x^{-1}
\dot m / M$. Then the emission measure of the wind reads
\begin{eqnarray}
n^2 V &=  & 2 \pi \, n_o^2 \; \frac{\dot m_o^2}{M^2} \; r_o^3 \,
M^3 \, x_m \; 2  \int_0^{\pi/2} A^2 \,\tilde n (\theta)^2 \, {\rm sin}
\theta \, d \theta \\
& \simeq & 10 \pi \, n_o^2 r_o^2 \sigma_T^2 \; \frac{
r_o}{\sigma_T^2} \; x_m \,\dot m_o^2 \, M \simeq 7.1 \times 10^{49}
r_o \, \dot m^2 \, M \, x_m
\end{eqnarray}
where the extra factor of 2 in front of the integral takes care of
the emission by both hemispheres, the numerical value of the
integral is $\simeq 2.5$, $\sigma_T$ is the Thomson cross section
and we have taken for the (normalized in units of $r_S$) maximum
radius of the disk the value $x_m \sim 10^6$, with $r_o \simeq r_S
\simeq 3 \times 10^5$ cm, the \sw radius of one solar mass. Then,
the H$\alpha$ line luminosity will be
\begin{equation}
L_{{\rm H}\alpha} \simeq \alpha(T)\, n^2  V \, \epsilon_{{\rm
H}\alpha} \simeq 2 \cdot 10^{36} \, \dot m_o^2 \, M ~{\rm erg
~s}^{-1}~.
\label{eq:lineL}
\end{equation}
At the same time, the accretion luminosity will be
\begin{equation}
L \simeq   10^{38} \; \eta \, \dot m_o \, M \label{eq:diskL}
\end{equation}
where $\eta \simeq 0.1$ is the accretion disk radiative efficiency
($\eta \simeq 0.3$ for neutron stars and $L \propto \dot m_o^2 \, M$
in the case of ADAF).

\begin{figure}[t]
\begin{center}$
\begin{array}{cc}
\includegraphics[trim=0in 0in 0in
0in,keepaspectratio=false,width=3.1in,angle=-0]
{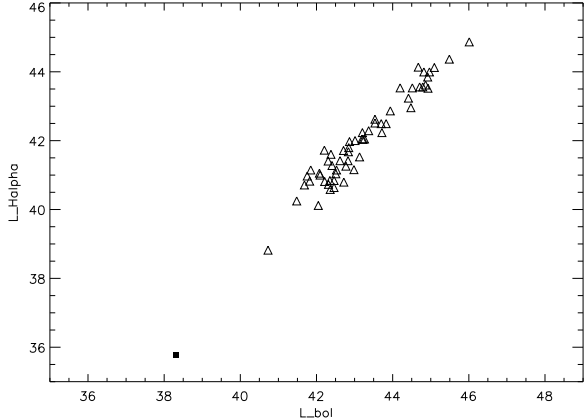} &
\end{array}$
\end{center}
\caption{\small The H$\alpha$ luminosity of a number of AGN (triangles) as
a function of their  bolometric luminosity (taken to be their
monochromatic luminosity at $\lambda = 4680$ \AA, as given in
Osterbrock 1989). The square denotes the H$\alpha$ luminosity of
Cir X-1 given in \citet{Johnston99} assuming the bolometric
luminosity to be that of Eddington for a 1.4 solar mass neutron
star. } \label{fig:fig6}
\end{figure}

The obvious point to note between Eqs (\ref{eq:lineL}) and
(\ref{eq:diskL}) is that they both are proportional to the object's
mass $M$; also for $\dot m_o < \alpha^2$ ($\alpha$ is the accretion
disk viscosity parameter) implying that the accretion flow is in the
ADAF regime, the ratio $R = L_{{\rm H}\alpha}/L$ of the line to the
bolometric accretion luminosity is independent of both $\dot m_o$
and the mass $M$. This ratio depends only on the recombination
coefficient $\alpha(T)({\rm with ~ assumed ~ value} \sim 10^{-13} ~
{\rm cm}^3 ~{s}^{-1}$ in Eq. (\ref{eq:lineL}) above) and the disk
efficiency $\eta$ and has a value $R \simeq 1/20 - 1/50$. The data
indicating the correlation between the H$\alpha$ line and the
continuum luminosity at $\lambda 4800$ \AA~is shown in Fig.
\ref{fig:fig6}. The data shown are those of 11.6 of \citet{Oster89}
with the abscissa converted from flux to luminosity; this figure
includes in addition the H$\alpha$ luminosity of Cir X-1
\citep[filled square;][]{Johnston99}, a galactic XRB; inclusion of
this point in the figure is important because it almost doubles the
dynamic range of this relation which now covers eight decades in the
line luminosity $L_{{\rm H}\alpha}$. The H$\alpha$ luminosity of
this source falls slightly below the extrapolation of a linear
relation from AGN to XRB. A likely reason for this could be the
reduced emissivity at small values of $x$ (small radii) due to the
higher gas temperature there. Incidentally, for winds with the
density profile of the \citet{BP82} scaling, i.e. $n(x) \propto
x^{-3/2}$, Eq. (\ref{eq:lineL}) would be proportional to ${\rm log}
x_m$ instead of $x_m$ and the line luminosity would be roughly
$10^5$ times smaller than observed.
We consider the straightforward, minimal assumption way that this
model accounts for the linear relation between the H$\alpha$ and
bolometric luminosities an indication of the validity of its
fundamental premises.

\section{Summary, Discussion}

\begin{figure}[t]
\begin{center}$
\begin{array}{cc}
\includegraphics[trim=0in 0in 0in
0in,keepaspectratio=false,width=3.1in,angle=-0]
{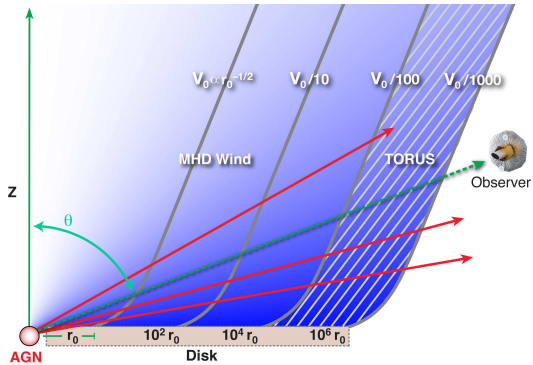} &
\end{array}$
\end{center}
\caption{\small A schematic of the model presented in \S 3.
The radius is shown in logarithmic scale, with the solid gray lines
representing the poloidal field at each radius shown and the
corresponding wind velocity relative to the fiducial one $v_o$.
The shading is proportional to the local column in the $r-$direction
which does not depend on $r$ but has a strong $\theta-$dependence
as required by AGN unification. At sufficiently large distances,
$r \gsim 10^4 \, r_o$ the flow is sufficiently cool to be molecular
and acts as the torus required by AGN unification.}
\label{fig:fig8}
\end{figure}

In the previous sections we presented a broad strokes picture of the
2D AGN structure, one that encompasses many decades in radius and
frequency and supplements the well known schematic of \citet{UP95}
with an outflow launched across the entire disk area and velocity
roughly equal to the local Keplerian velocity at each launch radius.
These outflows are responsible for the absorption features first
detected in the AGN UV spectra and more recently established also in
their X-ray ones.
As noted in the introduction, the goal and the spirit of this paper
is not to reproduce the detailed phenomenology  of specific AGN
spectral bands, but provide an account of their most general and
robust trends; however, it purports to do so on the basis of a
single, well founded, semi-analytic model for outflows off accretion
disks that involves a small number of free parameters. In this same
spirit, Fig. \ref{fig:fig8} presents a schematic of our model; the
radius is in logarithmic space and the shading is indicative of the
local {\sl column} density in the spherical$-r$ direction, which is
constant, but has a strong dependence on the angle $\theta$. The
gray lines are illustrative of the magnetic field geometry with the
corresponding scaling of the wind velocity on each one relative to
the fiducial one $v_o$ shown in the figure.

As shown in section \S 3, these outflows have the interesting
property that their ionization and dynamical structures scale mainly
with one parameter, the dimensionless accretion rate $\dot m$. As
such, they are applicable, in principle, to all accretion powered
sources from galactic accreting black holes to the most luminous
quasars. The mass of the object, $M$, simply provides the overall
scale of an object's luminosity and size (and also a characteristic
temperature for the BBB), in a way similar to that proposed by
\citet{Boroson02}.

However, despite this economy of parameters, because of the
inherently 2D character of these winds, their appearance (and that
of the AGN central regions) depends quite significantly on the
observer's inclination angle $\theta$, a desirable feature and in
agreement with our notions of AGN unification. Furthermore, and most
importantly, as noted in section \S 4, the ionization structure of
these flows depends also on the spectrum of the ionizing radiation.
This dependence complicates the situation because it breaks the
overall wind flow scale invariance on the mass $M$. As noted
earlier, it is effectively the dependence of $\alpha_{OX}$ on
luminosity which is responsible for the differences in the
absorption feature properties between Seyferts and BAL QSOs. This
dependence of $\alpha_{OX}$ on $L(2500$\AA) (and effectively on
$\dot M$) suggests that eventually the wind ionization properties
may constitute a two rather than three parameter family. Finally,
the high ionization of the inner regions of these flows in XRB,
naturally accounts for the low velocities of the Fe-K features
observed in galactic sources.

The crucial and fundamental aspect of the underlying MHD wind model
that allows the broad consolidation of the very diverse
observational phenomenology of the previous sections ``under the
same roof'' is their ability to produce density profiles that
decreases like $\sim 1/r$ with the radius. It is this property that
allows the ionization parameter to decrease with distance while
still providing sufficient column to allow the detection of both
high and low ionization ions in the AGN X-ray spectra. It is also
this property that controls the velocities $v$ of the Fe-K
transitions in galactic sources, Seyferts and BAL QSO, due to the
relation between $v$ and $\xi$ ($v^2 \propto \xi$ for the scaling
proposed herein; the BP82 scaling leads to $v \propto \xi$ and
therefore to much smaller velocity for a given ion, i.e. a given
value of $\xi$ -- and also to a much smaller column). Furthermore,
it is this specific density profile of the wind that allows us to
incorporate the physics of the AGN torii within the context of the
broader physics of accretion onto the compact object, while at the
same time producing IR spectra in broad agreement with observation.
This confluence of the AGN spectral properties with the wind spacial
structure indicates that AGN and XRB are objects that span many
decades in radius and frequency, despite the fact that most of their
luminosity is released within a few \sw radii.

The price to pay for a density distribution such as that proposed
above is the need to invoke winds whose mass flux increases
with distance from the source (and more specifically like $\sim
r^{1/2}$). While this feature can be accommodated by the choice of a
parameter in the models of \citet{CL94}, it was given a rather
transparent explanation in \citet{BB99}, in terms of the dynamics of
accretion. Interestingly, recent {\em Chandra} spectroscopy
\citep{BS00,Behar03,NRL} appear to support such a notion. At this
point, it is not obvious how theoretically compelling is this
particular feature in the general scheme of accretion properties.
Why is this mass loss preferred to one that would result in, for
instance,  the BP82 density profile? Are there AGN with winds/accretion
flows  consistent with $n(r) \propto r^{-3/2}$? What parameter determines
which one is chosen by nature? These are pressing questions for
which we currently have no answers.
However, the importance of the specific mass flux dependence on $r$
in the interpretation of the AGN X-ray and UV absorber properties
will likely attract  the attention of future studies on this issue.

Clearly a presentation as broad as that above by necessity ignores a
large number of issues, both observational and theoretical each of
which is in fact a separate branch in the study of AGN physics. As
such, we have ignored the effects of radiation pressure, considered
in much detail numerically by Proga and collaborators
\citep{PSK00,Proga03,Proga04} and semi-analytically by
\cite{MCGV95}. The effects of radiation pressure in combination with
those of the MHD winds discussed here have been considered by
\citet{Everett05} and also by \cite{KK94} in discussing the AGN IR
spectra. These works have shown the effects of radiation pressure to
be local, as they depend on the local flow opacity, thus breaking
the similarity of the angular dependence of the solutions. It was
shown in these works that radiation pressure ``pushes" to open up
the field lines to produce a density dependence on $\theta$
different from that given in Fig. \ref{fig:fig2}a and hence would
influence the population ratios of objects observed at a given
column density. However, the effects of radiation pressure do not
affect the dependence of wind mass flux on the distance ($\dot m
\propto r^{1/2}$), which is set by conditions on the disk; it is
this property that determines the radial dependence of the wind
density, the property necessary to account for their observed
phenomenology.

As noted above, a radiation driven wind, while 2D in the region of
launch, it will appear radial at sufficiently large distance
producing density profiles $\simeq 1/r^2$ there. As appealing and
compelling as the notion of radiation driven winds is, there is
little evidence for them, at least in the AGN and XRB X-ray absorber
spectra. Advocating radiation driven 2D winds across the entire
accretion disk, in a fashion similar to those of \S 3, appears
difficult because the photon field does not have enough momentum at
these large distances to drive a wind with the required mass flux.
Interestingly the $\phi-$component of the magnetic field, decreasing
like $B_{\phi} \propto 1/r$ has precisely the momentum needed to
drive a wind with $\dot m \propto r^{1/2}$. In this respect one
should bear in mind that these winds produce most of the kinetic
energy at small radii, most of the mass flux at large radii and
equal momentum per decade of radius.

Another issue that was only briefly touched upon in section 5, is
that of line emission. It is generally thought that the line
emission in AGN comes from clouds in pressure equilibrium with a hot
intercloud medium, the result of the X-ray heating thermal
instability \citep{Krolik81}. Our simple estimates, even though they
have ignored this possibility, they  nonetheless provide an account
for the observed correlation of the H$\alpha$ (a transition with
minimal radiative transfer nuances) with the AGN bolometric
luminosity. However, one should bear in mind that our model winds do
allow for the formation of such clouds at sufficiently small
latitudes (below the Alfv\'en surface) where the flow is close to
hydrostatic equilibrium, i.e. under conditions of a given pressure.
We expect that past the Alfv\'en point, where the flows are under
conditions of given density, the formation of these clouds will be
less forthcoming. The issue of cloud or wind AGN line emission is an
issue that deserves more attention and study, given the smoothness
of the AGN lines profiles that implies a very large number clouds
involved in this process \citep{Arav97}, but certainly beyond the
scope of the present paper.

Our treatment of the AGN IR emission has also glossed over much of
what it constitutes an altogether distinct subfield of AGN study. It
has been suggested that clouds are also involved in this component
of the AGN spectra, however with different properties and at radii
larger than those of the UV and optical line emitting clouds
\citep{Nenkova08}. This is clearly a point that will have to be
looked upon with greater care. A complicating factor in this
direction is that of star formation in the AGN environment, whose IR
contribution introduces additional parameters in such a study.
Finally, with respect to the IR and far-IR AGN spectra, we would
like to point to the synergy between X-ray spectroscopy and far-IR
observations discussed in \S 5.1 that would help establish the
consistency of this scheme across two very different frequency
bands.

Finally, we close with a few words on the ``feedback" of our winds
on the surrounding medium, which likely provides the mass that
eventually ``feeds" the AGN. First, the angular distribution of the
flow, as determined by the poloidal field structure, due to its
collimation, interacts with only part of the surrounding stellar
cluster. Second, the energy flux of the winds considered here,
despite their increasing mass flux with radius, is still dominated
by the flux at small radii ($\dot E = \dot m v^2/2 \propto
r^{-1/2}$); however, they do carry equal momentum per decade of
radius ($\dot P = \dot m v \propto r^0$), a fact with potential
feedback effects, perhaps the subject of a future publication.
Finally, the radiation effects of the AGN are also limited in
$\theta$ due precisely to the winds' column increase with this
parameter. However, as shown in \citet{Tueller08}, the fraction of
{\em Swift-BAT} selected AGN at galactic latitude $|b|>15^{\circ}$
with X-ray column $N_H > 10^{22} \, {\rm cm}^{-2}$ decreases from
$\simeq 0.5$ to close to zero above luminosity $L \simeq 10^{44} \;
{\rm erg \,s}^{-1}$. Therefore, considering that the observed column
pertains not only to the AGN wind but also to the entire matter
distribution along the LoS, the radiative effects of high luminosity
AGN may affect significantly the evolution of their environment.
Such constraints will have to be taken into account producing a
global evolutionary sequence for our models, which are beyond the
scope of the present paper.

%




\acknowledgments

Authors are grateful to Tim Kallman for insightful discussions and
help with running \verb'XSTAR'. K.F.
and D.K. would also like to thank G. Chartas, J. Turner, L. Miller,
S. Kraemer, T. Misawa for their constructive comments. D.K. would
like to thank R. Mushotzky, C. Reynolds and C. Miller for discussions and the
Astronomy Department of the University of Maryland for the hospitality
during his sabbatical visit there.

\end{document}